\documentclass{aa}  

\usepackage{mathabx} 
\usepackage{mathtools}
\usepackage{mathrsfs}
\usepackage{nicefrac}
\makeatletter
\newcommand{\ostar}{\mathbin{\mathpalette\make@circled\star}}
\newcommand{\make@circled}[2]{%
  \ooalign{$\m@th#1\smallbigcirc{#1}$\cr\hidewidth$\m@th#1#2$\hidewidth\cr}%
}
\newcommand{\smallbigcirc}[1]{%
  \vcenter{\hbox{\scalebox{0.77778}{$\m@th#1\bigcirc$}}}%
}

\usepackage{amsmath} 
\usepackage{array}
\usepackage{adjustbox}
\usepackage{multirow}
\usepackage{makecell}
\newcolumntype{C}[1]{>{\centering\arraybackslash}p{#1}} 
\newcolumntype{L}[1]{>{\raggedright\arraybackslash}m{#1}} 
\newcolumntype{M}[1]{>{\centering\arraybackslash}m{#1}} 
\newcolumntype{B}[1]{>{\centering\arraybackslash}b{#1}} 

\renewcommand{\refeq}[1]{Eq.~(\ref{#1})\xspace} 
\newcommand{\refeqs}[2]{Eqs.~(\ref{#1}) and (\ref{#2})\xspace}

\newcommand{\reffig}[1]{Fig.~\ref{#1}\xspace}
\newcommand{\reffigs}[2]{Figs.~\ref{#1} and~\ref{#2}\xspace} 
\newcommand{\reffigfull}[1]{Figure~\ref{#1}\xspace} 
\newcommand{\reffigsfull}[2]{Figures~\ref{#1} and~\ref{#2}\xspace} 

\newcommand{\refsub}[1]{#1} 
\newcommand{\refsubfig}[2]{Fig.~\ref{#1}\refsub{#2}\xspace}
\newcommand{\refsubfigfull}[2]{Figure~\ref{#1}\refsub{#2}\xspace} 
\newcommand{\refsubfigs}[2]{Figs.~\ref{#1}\refsub{#2}\xspace}

\newcommand{\refpan}[1]{Panel~#1} 
\newcommand{\refpans}[1]{Panels~#1} 

\newcommand{\reftab}[1]{Table~\ref{#1}\xspace}

\newcommand{\refalg}[1]{Algorithm~\ref{#1}\xspace}

\newcommand{\reflin}[2]{Line~\ref{lin:#1:#2}\xspace} 

\newcommand{\refsec}[1]{Sect.~\ref{#1}\xspace} 
\newcommand{\refsecs}[2]{Sects.~\ref{#1} and~\ref{#2}\xspace} 

\newcommand{\refapp}[1]{Appendix~\ref{#1}\xspace} 

\usepackage{graphicx}
\graphicspath{{./figures/}}

\usepackage{transparent}
\usepackage{xkeyval,xcolor}
\makeatletter
\newlength{\sfp@hseplen}\newlength{\sfp@vseplen}
\define@cmdkey{subfigpos}[sfp@]{pos}[ul]{}
\define@cmdkey{subfigpos}[sfp@]{font}[\small]{}
\define@cmdkey{subfigpos}[sfp@]{vsep}[0.75\baselineskip]{\setlength{\sfp@vseplen}{\sfp@vsep}}
\define@cmdkey{subfigpos}[sfp@]{hsep}[3.5pt]{\setlength{\sfp@hseplen}{\sfp@hsep}}
\newcommand{\subfigimg}[4][,]{%
	\setkeys{Gin,subfigpos}{pos,font,vsep,hsep,#1}
	\setbox1=\hbox{\includegraphics{#4}}
	\ifnum\pdfstrcmp{\sfp@pos}{ul}=0
		\leavevmode\rlap{\usebox1}
		\rlap{\hspace*{\sfp@hsep}\raisebox{\dimexpr\ht1-\sfp@vsep}{\transparent{#3}{\setlength{\fboxsep}{1pt}\colorbox{white}{%
\transparent{1}\sfp@font{#2}}}%
}}
		\phantom{\usebox1}
	\else\ifnum\pdfstrcmp{\sfp@pos}{ur}=0
		\leavevmode\usebox1
		\llap{\raisebox{\dimexpr\ht1-\sfp@vsep}{\sfp@font{#2}}\hspace*{\sfp@hsep}}
	\else\ifnum\pdfstrcmp{\sfp@pos}{lr}=0
		\leavevmode\usebox1
		\llap{\raisebox{\sfp@vsep}{\sfp@font{#2}}\hspace*{\sfp@hsep}}
	\else
		\leavevmode\rlap{\usebox1}
		\rlap{\hspace*{\sfp@hseplen}\raisebox{\sfp@vsep}{\sfp@font{#2}}}
		\phantom{\usebox1}
	\fi\fi\fi
}
\newcommand{\fontfig}[1]{\tiny$\!\!$\color{#1}\textbf}
\newcommand{\AspectRatio}[1]{\dimexpr 1pt * \wd#1 / \ht#1 \relax} 

\DeclarePairedDelimiterX{\paren}[1]{(}{)}{#1}
\newcommand{\Paren}[1]{\paren*{#1}}
\let\brace=\undefined 
\DeclarePairedDelimiterX{\brace}[1]{\{}{\}}{#1}
\newcommand{\Brace}[1]{\brace*{#1}}
\let\brack=\undefined 
\DeclarePairedDelimiterX{\brack}[1]{[}{]}{#1}
\newcommand{\Brack}[1]{\brack*{#1}}
\DeclarePairedDelimiterX{\bbrack}[1]{\llbracket}{\rrbracket}{#1}
\newcommand{\Bbrack}[1]{\bbrack*{#1}}
\DeclarePairedDelimiterX{\abs}[1]{\rvert}{\lvert}{#1}     
\newcommand{\Abs}[1]{\abs*{#1}}
\DeclarePairedDelimiterX{\norm}[1]{\lVert}{\rVert}{#1}    

\DeclarePairedDelimiterX{\avg}[1]{\langle}{\rangle}{#1}   

\DeclarePairedDelimiterX{\ceil}[1]{\lceil}{\rceil}{#1}     

\DeclarePairedDelimiterX{\floor}[1]{\lfloor}{\rfloor}{#1}  

\usepackage{algorithm}
\usepackage[noend]{algpseudocode}
\newcommand{\commentalgo}[1]{\Comment{{\tiny #1}}} 

\newcommand{\conv}{\star}				

\newcommand{\Tag}[1]{\text{#1}}			
\newcommand{\n}[1]{n^{\Tag{#1}}}		
\newcommand{\f}[2]{f^{\Tag{#1}}\Paren{#2}}		

\newcommand{\V}[1]{{\boldsymbol{#1}}}	

\newcommand{\data}{{d}}   		          	
\newcommand{\Vdata}{{\V{\data}}}          		
\newcommand{\obj}{{o}}   		          	
\newcommand{\Vobj}{{\V{\obj}}}           	
\newcommand{\x}{{x}}   		          	 	
\newcommand{\Vx}{{\V{\x}}}            	  	
\newcommand{\psf}{{p}}            			
\newcommand{\Vpsf}{{\V{\psf}}}            	
\newcommand{\pcore}{{\Vpsf^{\Tag{core}}}}   	
\newcommand{\weight}{{w}}            		
\newcommand{\Vweight}{{\V{w}}}            	
\newcommand{\weightrob}{\weight^{\rob}}

\newcommand{\regpar}{{\mu}}            		

\newcommand{\argmin}[2]{\underset{#1}{\text{argmin}}\;#2}
\newcommand{\mof}[1]{m\Paren{#1}}
\newcommand{\dat}[2]{\mathscr{D}^{\Tag{#1}}\Paren{#2}}
\newcommand{\cost}[3]{\mathscr{C}^{\Tag{#1}}_{#2}\Paren{#3}}
\newcommand{\reg}[2]{\mathscr{R}^{\Tag{#1}}\Paren{#2}}
\newcommand{\rob}{\rho}
\newcommand{\robdata}[1]{\mathscr{D}^{\rob}\Paren{#1}}
\newcommand{\throb}{\bar\weight^{\rob}}

\newcommand{\varRON}{v^{\Tag{ron}}}
\newcommand{\VvarRON}{\V{v}^{\Tag{ron}}}
\newcommand{\underseteq}[2]{\underset{\text{\refeq{#1}}}{#2}}

\newcommand{\st}{\text{ s.t. }\xspace}    

\usepackage[squaren,thickspace,thickqspace]{SIunits}
\newcommand{\percent}[1]{#1\,\%}     

\newcommand{\Kleopatra}{(216)~Kleopatra\xspace}
\newcommand{\Elektra}{(130)~Elektra\xspace}

\newcommand{\Nguu}{\includegraphics[height={\f@size pt*2/3}]{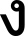}}
\newcommand{\Reua}{\includegraphics[height={\f@size pt*2/3}]{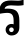}}
\newcommand{\Thong}{\includegraphics[height={\f@size pt*2/3}]{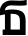}}

\usepackage{txfonts}
\usepackage[colorlinks=true, allcolors=blue]{hyperref}
\usepackage{flushend}
\usepackage{silence}
\begin{document}


\title{Blind and robust estimation of adaptive optics point spread function and diffuse halo with sharp-edged objects}
\subtitle{Application to asteroid deconvolution and moon enhancement\thanks{Based on observations made with ESO Telescopes at the Para{\-}nal Observatory under programmes ID 60.A-9362(A) (Yang et al. 2014), ID 296.C-5038(A) (Yang et al. 2016), ID 199.C-0074 (Vernazza et al. 2021) and reduced data publicly available at \href{https://observations.lam.fr/astero/}{https://observa-tions.lam.fr/astero/} and observations obtained at Keck Observatory under programme ID U58N2 (de~Pater et al. 2005).}}

\author{
	Anthony Berdeu\inst{1,2}
	}

\institute{
	LESIA, Observatoire de Paris, Université PSL, Sorbonne Université, Université Paris Cité, CNRS, 5 place Jules Janssen, 92195 Meudon, France
	\and
	National Astronomical Research Institute of Thailand, Center for Optics and Photonics, 260 Moo 4, T. Donkaew, A. Maerim, Chiang Mai 50180, Thailand
	\\
	\email{anthony.berdeu@obspm.fr}
}

\date{Received 2 August 2023 / Accepted 7 May 2024}

 
\abstract
{
Initially designed to detect and characterise exoplanets, extreme adaptive optics (AO) systems open a new window onto the Solar System by resolving its small bodies. Nonetheless, their study remains limited by the accuracy of the knowledge of the AO-corrected point spread function (AO-PSF) that degrades their image and produces a bright halo, potentially hiding faint moons in their close vicinity.
}
{
To overcome the random nature of AO-PSFs, I aim to develop a method that blindly recovers the PSF and its faint structured extensions directly into the data of interest, without any prior on the instrument or the object's shape. The objectives are both to deconvolve the object and to properly estimate and remove its surrounding halo to highlight potential faint companions.
}
{
My method first estimated the PSF core via a parametric model fit, under the assumption of a sharp-edged flat object. Then, the resolved object and the PSF extensions were alternatively deconvolved with a robust method, insensitive to model outliers, such as cosmic rays or unresolved moons. Finally, the complex halo produced by the AO system was modelled and removed from the data.
}
{
The method is validated on realistic simulations with an on-sky AO-PSF from the SPHERE/ZIMPOL instrument. On real data, the proposed blind deconvolution algorithm strongly improves the image sharpness and retrieves details on the surface of asteroids. In addition, their moons are visible in all tested epochs despite important variability in turbulence conditions.
}
{
My method shows the feasibility of retrieving the complex features of AO-PSFs directly from the data of interest. It paves the way towards more precise studies of asteroid surfaces and the discovery and characterisation of Solar System moons in archival data or with future instruments on extremely large telescopes with ever more complex AO-PSFs.
}

\keywords{instrumentation: adaptive optics - methods: numerical - techniques: high angular resolution – techniques: image processing - minor planets, asteroids: general}

\maketitle


\let\oldpageref\pageref
\renewcommand{\pageref}{\oldpageref*}

\section{Introduction}
\label{sec:intro}

The performance of ground-based instruments is limited by the atmospheric turbulence that corrugates the incident wavefront of the observed target~\citep{Roddier:81}. For short exposures, the diffraction-limited point spread function (PSF) of the telescope breaks down to a random speckle field. For long exposures, their average is equivalent to the PSF of a 10 to \unit{20}{\centi\meter} telescope~\citep{Fried:66_turbulence}, strongly degrading the sensitivity and the resolution.

Introduced in the 1990s, adaptive optics (AO) systems are now commonly deployed in instruments of ground-based observatories to mitigate the effects of the turbulence~\citep{Tyson:15_principles_of_AO}. The coupling of (i) PSF prediction to optimise the instrument design and performance~\citep{Fusco:06_SPHERE_design, Dohlen:16_SPHERE_performance_vs_prediction}, with (ii) PSF modelling~\citep{Jolissaint:06_AO_analytical, Fetick:19_model_based_AOPSF, Berdeu:23_IF_analytical} to improve the data post-processing via model-fitting or deconvolution~\citep{Beltramo:20_PSF_reconstruction_Review}, has resulted in a leap forward in numerous fields such as photometry and astrometry of stellar populations~\citep{Turri:17_astrometry_MCAO, Monty:18_astrometry_cluster}, spectroscopy and kinematics of distant galaxies~\citep{Schreiber:18_SINFONI, Bianchin:22_Gemini_NIFS}, or the study of extended object surface topology~\citep{Rimmele:21_solar_AO, Vernazza:21_Large_program}.

In the last decade, the arrival of extreme AO systems pushed the performance in high-contrast and high-resolution imaging even further ~\citep{Jovanovic:15_XAO}. In this context, PSF prediction and modelling reach their limits, unable to faithfully reproduce the complexity of observed AO-corrected PSFs (AO-PSFs). Point spread function estimation or reconstruction techniques must then be considered~\citep{Beltramo:20_PSF_reconstruction_Review}. But direct estimation of the PSF parameters from the AO telemetry~\citep{Veran:97_PSF_AO_telemetry, Clenet:08_NACO_PSF_recons} or reference PSFs obtained on calibration sources (internal or natural stars) before or after the observation~\citep{Mugnier:04_Mistral} are not always sufficient, and can for example lead to strong deconvolution artefacts~\citep{Marchis:06_moon_detection, Fetick:19_Vesta, Fetick:20_param_marignal, Lau:23_prior_AOPSF}.

Among others, the challenge arises from the evolving nature of AO-PSFs. By essence, the turbulence is random and the AO loops are not perfect, suffering from measurement noise, or temporal and aliasing errors~\citep{Rigaut:98_SH_error}. Finally, flexions or thermal dilations in the instruments induce quasi-static speckles that slowly drift with time, changing the PSF's small and faint structures~\citep{Milli:2016_quasi_static_XAO, Vigan:19_ZELDA}. As a consequence, a given observation represents one realisation of an AO-PSF that is barely reproducible, or even impossible to reproduce.

To overcome this random nature of the AO-PSF, the only solution is to extract and reconstruct it directly from the data of interest, a problem known as blind deconvolution~\citep{Stockham:75_blind_deconvolution, Thiebaut:95_blind_deconvolution, Soulez:12, Fetick:20_param_marignal}. The challenge is to overcome the degeneracy between the estimation of the PSF and the object~\citep{Little:83_joint_estimation, Blanco:11_blind_retina} and to solve an ill-posed problem with potentially twice more unknowns than measurements if both the images of the PSF and the object are to be retrieved from a single data image. As a consequence, no method is strictly `blind', and there is a need to make some assumptions about the PSF or the deconvolved object (or both) to avoid local minima of the problem and converge on the best `PSF and object' pair.

One solution is to implement marginal approaches \citep{Blanc:03_Marginal_Zernike, Beltramo:20_PSF_reconstruction_Review}. The problem is rewritten to split the PSF and the object contributions, in order to integrate over the space of the object parameters to marginalise them out of the problem. The deconvolution of the object is only performed as a second step. Marginalisation has proved to be efficient when combined with parametric PSF models that strongly limit the number of unknowns to fit~\citep{Fetick:19_model_based_AOPSF}. These methods nonetheless imply a need to know some prior on the object, such as the structure of its power spectrum density  \citep{Fetick:20_param_marignal, Yan:23_myopic_MCMC} or its overall shape \citep{Lau:23_prior_AOPSF}. In addition, such PSF models, relying on a limited number of parameters, cannot fully grasp the complexity of real and potentially broadband AO-PSFs of high-contrast and high-resolution imagers. The challenge will be even greater with the incoming extremely large telescopes~(ELTs). Their segmented primary mirror and large spiders holding the secondary mirror will produce intricate and structured PSFs that are very sensitive to AO residuals~\citep{Hippler:19_METIS, Neichel:20_TIPTOP, Simioni:20_MICADO, Hedglen:22_GMT}. More versatile methods than parametric approaches are consequently needed.

Finally, in the context of asteroid study, beyond the deconvolution of the images of asteroids' surfaces~\citep{Vernazza:21_Large_program}, the approximate knowledge of the PSF also limits the study of asteroids' close vicinity and the detection of faint moons. Similar constraints apply when observing the moon systems around the giant planets of our Solar System \citep{Showalter:06_Uranus, Assafin:08_digital_coronography}. To get rid of the bright halo induced by the PSF extensions around the main object, techniques were adapted from exoplanet detection algorithms, based on local averaging or median filters~\citep{Marchis:06_moon_detection, Assafin:08_digital_coronography, Pajuelo:18_Carry_coronagraph}. But the problem to solve is quite different, the halo being produced not by a coherent source (an unresolved star) but by an extended object (the resolved asteroid or planet). These techniques consequently do not account for the physics at the origin of the halo that is poorly estimated. They thus inherently suffer from self-subtraction problems and can bias the moon photometry or even prevent its detection~\citep{Yang:16_Elektra_Minerva}.

It is thus essential to properly recover the AO-PSF extensions. But compared to the PSF core, their intensity is several orders of magnitude fainter. This echoes high-contrast imaging challenges where outliers must be carefully handled to avoid any corruption of the model. Some outliers, such as dead pixels, can be identified through proper calibration~\citep{Berdeu:20_PIC}, handled, and removed from the problem. But others may be random or specific to a given acquisition, such as hot pixels, cosmic rays, or model errors (signals not predicted by the forward model of the problem). This second kind of outlier must be identified on the fly, directly in the data of interest.

To reduce the incidence of such outliers, robust penalisation approaches have been developed~\citep[see][Chap. 5, for detailed overviews]{Zoubir:18_robust_stat, Flasseur:19_PhD}. They consist of replacing the conventional quadratic penalisation of the problem by a robust estimator that is approximately quadratic around zero but that grows sub-quadratically for large deviations to reduce their impact~\citep{Hogg:79, Huber:96}.

One solution is to directly change the cost function of the problem~\citep{Yohai:87_Robust_regression, Huber:08_High_Breakdown_robust, Huber:11_robust}. Another solution, so-called iterative reweighted least squares \citep[IRLS,][]{Holland:77, Sigl:16_nonlinear_IRLS}, is to stay in the least squares framework but to solve a sequence of least squares problems whose weights are iteratively updated with a robust estimator. The mathematical formalism associated with such robust penalisations will be further detailed in \refsec{sec:wing}.

This paper is a continuation of a previous pragmatic method that was initially developed to detect and characterise the third moon of \Elektra~\citep{Berdeu:22_Elektra}. In this previous work, the performance of an AO system being better in the infrared, a simple parametric model was sufficient to describe the PSF, to deconvolve the object, and to properly model and subtract its bright halo. In addition, the PSF was directly estimated on a bright moon in the data of interest, acting as a reference point source.

In this work, I intend to provide a more general and blind approach for more complex AO-PSFs with limited priors on the object and the PSF, while removing the need for a point source to be present in the field of view (FoV). There are three main objectives: (i)~to reconstruct the AO-PSF in its complexity, reproducing its faint structured extensions, so-called wings, directly from the data; (ii)~to deconvolve the object image with this blindly estimated AO-PSF; and (iii)~to carefully model the AO residual halo in order to remove it from the data and enhance the signal of potential moons.

The proposed method, which alternates between object and PSF deconvolution, is described in \refsec{sec:method}. It is then validated on realistic simulated data in \refsec{sec:res_sim} and applied to real data in \refsec{sec:res_real}. Finally, in 
\refsec{sec:conclu}, I discuss several ways of further improving the proposed method. For readers interested in knowing more about the algorithms and their implementation, the pseudo-codes of the different steps are detailed in \refapp{app:algo}.

\section{Proposed method}
\label{sec:method}

\begin{figure*}[t!]
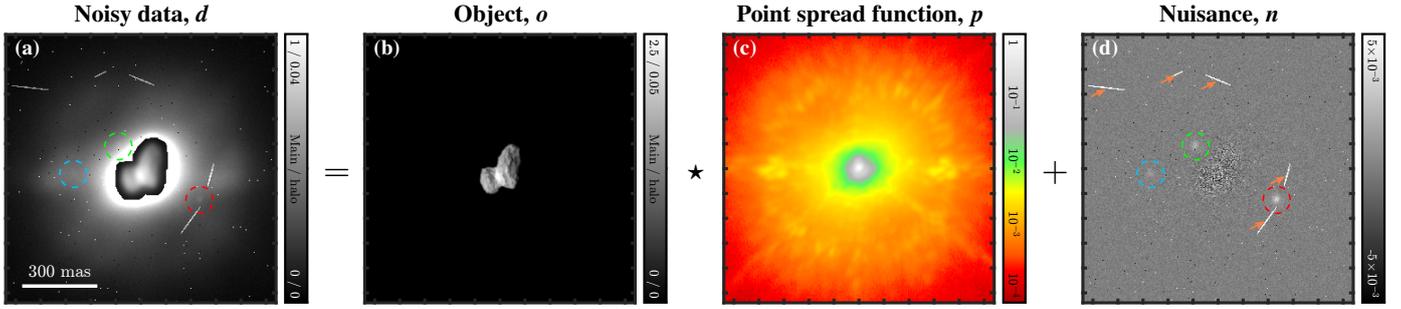
 
        \centering
        
        \newcommand{\PathFig}{Fig_overview/}
        \newcommand{\FigSuf}{_sim}
        
        \newcommand{\LineRatio}{0.95}
        
        \newcommand{\fontTxt}[1]{\textbf{\small #1}}
        
        \newcommand{\widthTxt}{16pt}
        
        \newcommand{\sizeTxt}[1]{\Large{#1}}
        
        \newcommand{\spaceLine}{1.8cm}
        
        \newcommand{\widthFig}{\dimexpr (\linewidth - \widthTxt * 3)}
        
        \newcommand{\subfigColor}{white}

        \newcommand{\FigOne}{\PathFig Data\FigSuf}
        \newcommand{\FigTwo}{\PathFig Obj\FigSuf}
        \newcommand{\FigThree}{\PathFig PSF\FigSuf}
        \newcommand{\FigFour}{\PathFig Res\FigSuf}
        \newcommand{\FigOneBar}{\PathFig Data_bar}
        \newcommand{\FigTwoBar}{\PathFig Obj_bar}
        \newcommand{\FigThreeBar}{\PathFig PSF_bar}
        \newcommand{\FigFourBar}{\PathFig Res_bar}
        
        \sbox1{\includegraphics{\FigOne}}
        \sbox2{\includegraphics{\FigOneBar}}
        \sbox3{\includegraphics{\FigTwo}}
        \sbox4{\includegraphics{\FigTwoBar}}
        \sbox5{\includegraphics{\FigThree}}
        \sbox6{\includegraphics{\FigThreeBar}}
        \sbox7{\includegraphics{\FigFour}}
        \sbox8{\includegraphics{\FigFourBar}}
        
        \newcommand{\ColumnWidth}[1]
                {\dimexpr \LineRatio \widthFig * \AspectRatio{#1} / (\AspectRatio{1} + \AspectRatio{2} + \AspectRatio{3} + \AspectRatio{4} + \AspectRatio{5} + \AspectRatio{6} + \AspectRatio{7} + \AspectRatio{8}) \relax
                }
        \newcommand{\ColumnGap}{\hspace {\dimexpr \widthFig /13 - \LineRatio\widthFig /13 }}

        \begin{tabular}{
                @{\ColumnGap}
                M{\ColumnWidth{1}}
                @{\ColumnGap}
                M{\ColumnWidth{2}}
                @{\ColumnGap}
                M{\widthTxt}
                @{\ColumnGap}
                M{\ColumnWidth{3}}
                @{\ColumnGap}
                M{\ColumnWidth{4}}
                @{\ColumnGap}
                M{\widthTxt}
                @{\ColumnGap}
                M{\ColumnWidth{5}}
                @{\ColumnGap}
                M{\ColumnWidth{6}}
                @{\ColumnGap}
                M{\widthTxt}
                @{\ColumnGap}
                M{\ColumnWidth{7}}
                @{\ColumnGap}
                M{\ColumnWidth{8}}
                @{\ColumnGap}
                }
                
                \fontTxt{Noisy data, $\Vdata$}& &
                &
                \fontTxt{Object, $\Vobj$}& &
                &
                \fontTxt{Point spread function,  $\Vpsf$}& &
                &
                \fontTxt{Nuisance,  $\V{n}$} &
                \\                
                
                \subfigimg[width=\linewidth,pos=ul,font=\fontfig{\subfigColor}]{$\;$(a)}{0.0}{\FigOne} &
                \subfigimg[width=\linewidth,pos=ul,font=\fontfig{\subfigColor}]{}{0.0}{\FigOneBar} &
                \sizeTxt{$=$} &
                \subfigimg[width=\linewidth,pos=ul,font=\fontfig{\subfigColor}]{$\;$(b)}{0.0}{\FigTwo} &
                \subfigimg[width=\linewidth,pos=ul,font=\fontfig{\subfigColor}]{}{0.0}{\FigTwoBar} &
                \sizeTxt{$\conv$} &
                \subfigimg[width=\linewidth,pos=ul,font=\fontfig{\subfigColor}]{$\;$(c)}{0.0}{\FigThree} &
                \subfigimg[width=\linewidth,pos=ul,font=\fontfig{\subfigColor}]{}{0.0}{\FigThreeBar} &
                \sizeTxt{$+$} &
                \subfigimg[width=\linewidth,pos=ul,font=\fontfig{\subfigColor}]{$\;$(d)}{0.0}{\FigFour} &
                \subfigimg[width=\linewidth,pos=ul,font=\fontfig{\subfigColor}]{}{0.0}{\FigFourBar}
        \end{tabular}        
        \caption{\label{fig:deconv_model} Simulation of the forward model of the deconvolution. The noisy data,~$\Vdata$~(\refpan{a}), is the convolution of an extended object,~$\Vobj$~(\refpan{b}), with the AO-corrected PSF of the instrument,~$\Vpsf$~(\refpan{c}), plus a nuisance term,~$\V{n}$~(\refpan{d}). This term is composed of the acquisition noises (readout + photon), defective pixels (`salt \&{} pepper' pattern), cosmic ray impacts (orange arrows), and signals from potential moons orbiting the main object (coloured circles). \refpans{a,b}: To emphasise both the main body and the surrounding halo, a dual linear scale was used to insert the main body in its halo, as is noted by the `/' in the colour bars. \refpans{b}: Photo of 67P/Churyumov–Gerasimenko by Rosetta (European Space Agency). \refpans{c}: PSF obtained with ZIMPOL on HD16469 (P.~Vernazza), normalised to peak at one for the display.}
\end{figure*}

In the isoplanetic domain (the PSF does not depend on the position in the FoV), the forward model of the problem  canonically states that the image data,~$\Vdata$ (discretised onto the sensor pixel grid), is the (discrete) convolution~$\conv$ of the extended object,~$\Vobj$, with the long exposure PSF,~$\Vpsf$, that combines the telescope and instrument response and the AO residuals~\citep{Fetick:20_param_marignal},
\begin{equation}
	\label{eq:forward_model}
	\Vdata = \Paren{\Vobj \conv \Vpsf} + \V{n}
	\,,
\end{equation}
where~$\V{n}$ is a nuisance term. This equation is pictured in \reffig{fig:deconv_model}. As is seen in \refsubfig{fig:deconv_model}{d}, the nuisance term encompasses the classical detector readout noise and photon shot noise, but also outliers such as defective pixels (`salt \& pepper' noise) and cosmic rays that may hit the sensor during the acquisition (orange arrows). I emphasise here that the way the problem is posed -- the convolution of an extended object with a PSF -- implies that the unresolved moons (point sources), are also considered to be outliers, and thus belong to the nuisance term (coloured circles). As was stated above, the objective is to split~$\Vobj$ and~$\Vpsf$ only from the knowledge of~$\Vdata$ without being corrupted by~$\V{n}$.

The PSF presented in \refsubfig{fig:deconv_model}{c} was obtained on a star, as part of the European Southern Observatory (ESO) Large Programme ID 199.C-0074~\citep[PI:][]{Vernazza:21_Large_program} and using the imaging mode of the Zurich IMaging POLarimeter instrument~\citep[ZIMPOL,][]{Schmid:18_ZIMPOL}. This figure emphasises all the complexity of an AO-PSF. The AO cut-off frequency (orange annulus) delimits the classical two regimes of AO-PSFs: (i)~the outer region (red), dominated by the atmospheric turbulence halo and left uncorrected by the AO system, and (ii)~the inner region dominated by the AO residuals. The roughly symmetric core of the PSF (white and grey) is surrounded by a wind-driven halo (green) elongated in the wind direction (temporal error of the AO system, so called servo-lag) and a speckle field produced by the non-common path aberrations~\citep[NCPAs,][]{Vigan:19_ZELDA} between the science part of the instrument and the AO system (yellow and orange). The diffraction spikes produced by the spiders holding the secondary mirror are also visible (yellow cross along the diagonals). All these features are responsible for the extended and structured bright halo visible in \refsubfig{fig:deconv_model}{a} that hides the neighbouring moons.

I propose to solve \refeq{eq:forward_model} step by step, using minimal assumptions on the object and the PSF, summarised in \reftab{tab:deconv_assumtions}. (I) Firstly, as is pictured in \reffig{fig:deconv_core}, the problem can be approximated, `from a distance', as a sharp-edged flat object, \refsubfig{fig:deconv_core}{b}, convolved with a simple PSF core, \refsubfig{fig:deconv_core}{c}. In the following, such an object will qualify as `binary' in the sense of morphological operations~\citep{Gonzalez:20_Matlab}. This strongly simplifies the problem in order to get a first estimated separation between the object,~$\Vobj$, and the PSF,~$\Vpsf$. (II) Secondly, as is pictured in \reffig{fig:deconv_obj}, this PSF core was used to deconvolve the main extended object, \refsubfig{fig:deconv_obj}{b}. (III) Thirdly, as is pictured in \reffig{fig:deconv_wing}, the deconvolution paradigm is reversed and the object was used to deconvolve the faint PSF extensions, so-called wings, \refsubfig{fig:deconv_wing}{c}. Finally, after the removal of the halo model, \refsubfig{fig:deconv_wing}{a}, the moon can be seen in the residuals, \refsubfig{fig:deconv_wing}{d}. A detection algorithm could then be applied to find the signal of potential faint moons further hiding in the noise. This is nonetheless beyond the scope of this paper. All of these steps are further detailed in the following.

\begin{table}[t!] 
    \caption{\label{tab:deconv_assumtions} Implied assumptions to perform the blind deconvolution.}
    \centering
    \begin{tabular}{cc}
    \hline
    \hline
    Object & PSF
    \\
    \hline
    \makecell{
    -- Extended and resolved [1, 2] \\
    -- Approximately flat [1] \\
    -- Smooth shape [1] \\
    -- Sharp edges [2, 3] \\
    -- Positive [2]
    }
    &
    \makecell{
    -- Core approximately \\
    described by a \\
    parametric model [1] \\
    -- Smooth structures [3] \\
    -- Positive [3]
    }
    \\
    \hline
    \end{tabular}
    \tablefoot{The numbers in the brackets indicate the steps of the method for which the assumptions are used.}
\end{table}

\subsection{Step 1. Estimation of the PSF core}
\label{sec:core}

\begin{figure*}[t!]
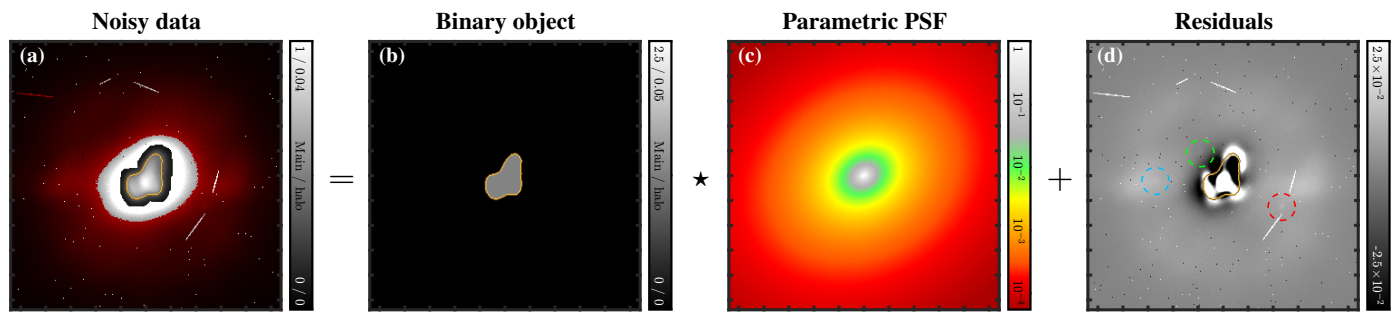
 
        \centering
        
        \newcommand{\PathFig}{Fig_overview/}
        \newcommand{\FigSuf}{_core}

        \newcommand{\LineRatio}{0.95}
        
        \newcommand{\fontTxt}[1]{\textbf{\small #1}}
        
        \newcommand{\widthTxt}{16pt}
        
        \newcommand{\sizeTxt}[1]{\Large{#1}}
        
        \newcommand{\spaceLine}{1.8cm}
        
        \newcommand{\widthFig}{\dimexpr (\linewidth - \widthTxt * 3)}
        
        \newcommand{\subfigColor}{white}

        \newcommand{\FigOne}{\PathFig Data\FigSuf}
        \newcommand{\FigTwo}{\PathFig Obj\FigSuf}
        \newcommand{\FigThree}{\PathFig PSF\FigSuf}
        \newcommand{\FigFour}{\PathFig Res\FigSuf}
        \newcommand{\FigOneBar}{\PathFig Data_bar}
        \newcommand{\FigTwoBar}{\PathFig Obj_bar}
        \newcommand{\FigThreeBar}{\PathFig PSF_bar}
        \newcommand{\FigFourBar}{\PathFig Res_core_bar}
        
        \sbox1{\includegraphics{\FigOne}}
        \sbox2{\includegraphics{\FigOneBar}}
        \sbox3{\includegraphics{\FigTwo}}
        \sbox4{\includegraphics{\FigTwoBar}}
        \sbox5{\includegraphics{\FigThree}}
        \sbox6{\includegraphics{\FigThreeBar}}
        \sbox7{\includegraphics{\FigFour}}
        \sbox8{\includegraphics{\FigFourBar}}
        
        \newcommand{\ColumnWidth}[1]
                {\dimexpr \LineRatio \widthFig * \AspectRatio{#1} / (\AspectRatio{1} + \AspectRatio{2} + \AspectRatio{3} + \AspectRatio{4} + \AspectRatio{5} + \AspectRatio{6} + \AspectRatio{7} + \AspectRatio{8}) \relax
                }
        \newcommand{\ColumnGap}{\hspace {\dimexpr \widthFig /13 - \LineRatio\widthFig /13 }}

        \begin{tabular}{
                @{\ColumnGap}
                M{\ColumnWidth{1}}
                @{\ColumnGap}
                M{\ColumnWidth{2}}
                @{\ColumnGap}
                M{\widthTxt}
                @{\ColumnGap}
                M{\ColumnWidth{3}}
                @{\ColumnGap}
                M{\ColumnWidth{4}}
                @{\ColumnGap}
                M{\widthTxt}
                @{\ColumnGap}
                M{\ColumnWidth{5}}
                @{\ColumnGap}
                M{\ColumnWidth{6}}
                @{\ColumnGap}
                M{\widthTxt}
                @{\ColumnGap}
                M{\ColumnWidth{7}}
                @{\ColumnGap}
                M{\ColumnWidth{8}}
                @{\ColumnGap}
                }
                
                \fontTxt{Noisy data}&&
                &
                \fontTxt{Binary object}&&
                &
                \fontTxt{Parametric PSF}&&
                &
                \fontTxt{Residuals}
                \\                
                
                \subfigimg[width=\linewidth,pos=ul,font=\fontfig{\subfigColor}]{$\;$(a)}{0.0}{\FigOne} &
                \subfigimg[width=\linewidth,pos=ul,font=\fontfig{\subfigColor}]{}{0.0}{\FigOneBar} &
                \sizeTxt{$=$} &
                \subfigimg[width=\linewidth,pos=ul,font=\fontfig{\subfigColor}]{$\;$(b)}{0.0}{\FigTwo} &
                \subfigimg[width=\linewidth,pos=ul,font=\fontfig{\subfigColor}]{}{0.0}{\FigTwoBar} &
                \sizeTxt{$\conv$} &
                \subfigimg[width=\linewidth,pos=ul,font=\fontfig{\subfigColor}]{$\;$(c)}{0.0}{\FigThree} &
                \subfigimg[width=\linewidth,pos=ul,font=\fontfig{\subfigColor}]{}{0.0}{\FigThreeBar} &
                \sizeTxt{$+$} &
                \subfigimg[width=\linewidth,pos=ul,font=\fontfig{\subfigColor}]{$\;$(d)}{0.0}{\FigFour} &
                \subfigimg[width=\linewidth,pos=ul,font=\fontfig{\subfigColor}]{}{0.0}{\FigFourBar}
        \end{tabular}
                
        \caption{\label{fig:deconv_core} Step~1 -- Estimation of the PSF core. The data (\refpan{a}, red-shaded pixels excluded from the fit) is approximated by the convolution of a binary object~(\refpan{b}, contour highlighted in orange) with a parametric PSF core~(\refpan{c}). \refpan{d}: residuals. Colour bars: see \reffig{fig:deconv_model}.
        }
\end{figure*}

The main objective of this step is to estimate a first separation between the object,~$\Vobj$, and the PSF,~$\Vpsf$, as is shown in \reffig{fig:deconv_core}. As a consequence, it must depend on a very limited number of parameters to be robustly constrained by the data.

Concerning the PSF, the method assumes that its core can be approximated by a parametric description. Such a simple model does not include the physics of an AO system and cannot properly model the turbulent halo. It thus dilutes the energy and is consequently not quantitative~\citep{Fetick:20_param_marignal}. Parametric model-based profiles can partially solve this issue \citep{Fetick:19_model_based_AOPSF} but they imply some knowledge of the instrument (such as, for example, its pupil shape, the number of actuators of its AO system, or the wavelength). As a consequence, they are harder to tune. But such precision is not needed for this given step. Indeed, \cite{Fetick:19_Vesta} showed that a Moffat profile~\citep{Moffat:69} is a general but sufficient approximation to qualitatively retrieve the morphology of the observed objects, while depending only on a limited number of parameters. A two-dimensional (2D) Moffat pattern is defined as
\begin{equation}
	\label{eq:Moffat}
    \mof{\Vx, \V{\alpha}, \beta, \theta} = 
    	\Paren{1 + 
        r_{1}^{2}/\alpha_{1}^{2}
        + r_{2}^{2}/\alpha_{2}^{2}
        }^{-\beta}
	\,,
\end{equation}
where $\beta$ is the power parameter of the pattern, $\theta$ is its orientation, $r_{1} = \x_{1}\cos{\theta} + \x_{2}\sin{\theta}$ and $r_{2} = -\x_{1}\sin{\theta} + \x_{2}\cos{\theta}$ are the 2D coordinates in its rotated frame, and $\V{\alpha} = \Paren{\alpha_{1}, \alpha_{2}}$ are its elongations along its two axes.

As we shall see later in \refsecs{sec:res_sim}{sec:res_real}, this Moffat approximation worked well with all the tested datasets. Nonetheless, in the hypothetical case that this model is not sufficient for a given dataset, \refeq{eq:Moffat} can be replaced by another more adapted parametric model if needed. Doing so does not change the presented methodology as long as the number of parameters remains limited.

The PSF core is thus simply proportional to this 2D pattern:
\begin{equation}
	\label{eq:PSF_core}
	\pcore\Paren{\Vx_{0}, \V{\alpha}, \beta, \theta, \gamma}
		{}\triangleq{}
		\gamma\mof{\Vx-\Vx_{0}, \V{\alpha}, \beta, \theta}
	\,.
\end{equation}
Its parameters mainly control how the object is blurred by the PSF. Most of the signal is then given by the object's edges rather than its surface texture. If the shape of the object is overall smooth with sharp edges, applying a threshold,~$\bar{d}$, on the data,~$\Vdata$, defined as
\begin{equation} 
	\label{eq:thres}
    \Brack{\f{thr}{\Vdata, \bar{d}}}\Paren{\Vx} = 
    \begin{cases}
		1 \text{ if } \data\Paren{\Vx} \geq \bar{d}
        \\
        0 \text{ otherwise}
	\end{cases}
	\,,
\end{equation}
is a sufficient approximation of its support to extract the information needed to estimate the parameters of \refeq{eq:PSF_core}. Indeed, in a parametric approach with a limited number of arguments, namely the parameters of the core and the threshold on the data, a crude description of the object is sufficient because there are far more measurements than unknowns to recover.

The values of the different parameters of \refeqs{eq:PSF_core}{eq:thres} are obtained by minimising the following cost function,
\begin{equation}
	\label{eq:cost_core}
	\begin{aligned}
		\cost{core}{\Vdata, \V{w}}{\bar{d}, \gamma, \Vx_{0}, \V{\alpha}, \beta, \theta} {}\triangleq{} &
		\\
		& \hspace{-35pt} \dat{wls}{
			\Vdata,
			\pcore\Paren{\Vx_{0}, \V{\alpha}, \beta, \theta, \gamma}\conv\f{thr}{\Vdata, \bar{d}},
			\Vweight
		}
		\,,
	\end{aligned}
\end{equation}
which is a pure data fidelity term between the data,~$\Vdata$, and the model, $\pcore\Paren{\Vx_{0}, \V{\alpha}, \beta, \theta, \gamma}\conv\f{thr}{\Vdata, \bar{d}}$, weighted by~$\Vweight$. $\mathscr{D}^{\Tag{wls}}$ is defined as the weighted least square difference (wls)
\begin{equation}
	\label{eq:WLS_full}
	\begin{aligned}
		\dat{wls}{\V{\varphi}_{1},\V{\varphi}_{2},\Vweight}
		{}\triangleq{} &
		\\
		& \hspace{-35pt}
			\frac{1}{2}\sum_{\Vx, \Vx^{\prime}} \weight\Paren{\Vx, \Vx^{\prime}}
		\Paren{\varphi_{1}\Paren{\Vx} -\varphi_{2}\Paren{\Vx}}
		\Paren{\varphi_{1}\Paren{\Vx^{\prime}} -\varphi_{2}\Paren{\Vx^{\prime}}}
		\,,
	\end{aligned}
\end{equation}
where~$\Vweight$ is the inverse of the data covariance matrix. Its role is to weigh the measurements in the data fidelity term in terms of confidence.

The readout noise is assumed to be of a Gaussian statistic of variance,~$\VvarRON$. On the other hand, for fluxes higher than a few photons, the Poisson statistic of photon shot noise can be approximated by a Gaussian statistic whose variance is proportional to the incident flux (and thus approximately to the data after proper calibration and pre-reduction) by a factor,~$\V{\eta}$, that accounts for the conversion between photons and analogue to digital unit (ADU) and for the quantum efficiency of the pixels~\citep{Berdeu:20_PIC}. In the general case, these noise terms are independent from one pixel of the sensor to another. As a consequence, the inverse of the data covariance is diagonal,
\begin{equation}
	\label{eq:weight_diag}
	\weight\Paren{\Vx, \Vx^{\prime}} = 
    \begin{cases}
		\weight\Paren{\Vx} \text{ if } \Vx = \Vx^{\prime}
        \\
        0 \text{ otherwise}
	\end{cases} 
	\,,
\end{equation}
and is given by~\citep{Mugnier:04_Mistral, Fetick:19_model_based_AOPSF}
\begin{equation}
	\label{eq:weight}
	\weight\Paren{\Vx} = 1/\Paren{\eta\Paren{\Vx}\data\Paren{\Vx}+\varRON\Paren{\Vx}}
	\,.
\end{equation}
The sum in \refeq{eq:WLS_full} is thus a sum on~$\Vx$ only
\begin{equation}
	\label{eq:WLS}
	\dat{wls}{\V{\varphi}_{1},\V{\varphi}_{2},\Vweight}
	{}\triangleq{}
	\frac{1}{2}\sum_{\Vx} \weight\Paren{\Vx}
	\Paren{\varphi_{1}\Paren{\Vx} -\varphi_{2}\Paren{\Vx}}^{2}
	\,.
\end{equation}
I emphasise that~$\V{\eta}$ and ~$\V{\varRON}$ are technically vectors as their values may depend on the position in the FoV, for example because of inhomogeneities in the pixel response or detector gains. Thus, contrary to other methods that approximate the noise level as being homogeneous across the pixels, to allow the marginalisation of the object and the PSF in the Fourier domain~\citep{Lau:23_prior_AOPSF, Yan:23_myopic_MCMC}, the proposed method includes a physical model of the noise statistics that depends on the pixels and that is a function of the data intensities (measured, see \refeq{eq:weight_diag}, or expected by the model, see \refeq{eq:weight_mod} below). This subtlety is critical: for the other methods, the unknown is the deconvolved image of the bright asteroid for which a homogeneous photon noise is a good enough approximation. But the objectives of my method are to additionally retrieve the PSF wings and remove the diffuse halo, while discarding faint outliers, which implies having a fine model of the noise according the intensity level.

As was discussed above, most of the signal is given by the blurring of the object's edges, and is condensed in the bright core of the halo. To avoid any corruption by the faint extensions of the halo that are not modelled under the parametric PSF core assumption, a threshold is applied to the data to reweight,~$\Vweight$, as is detailed in the \refalg{alg:core} of \refapp{app:algo_core}. Only the pixels not shaded in red in \refsubfig{fig:deconv_core}{a} are considered in the fit.

Finally, I mention that this first step can be replaced by more classical approaches, for example using reference PSFs obtained on a star or myopic approaches \citep{Mugnier:04_Mistral}. Indeed, its main objective is to get a good enough PSF core to initialise the object deconvolution of \refsec{sec:deconv}. The global PSF will be further refined later, in \refsec{sec:wing}.

\subsection{Step 2. Deconvolution of the main object}
\label{sec:deconv}

\begin{figure*}[t!]
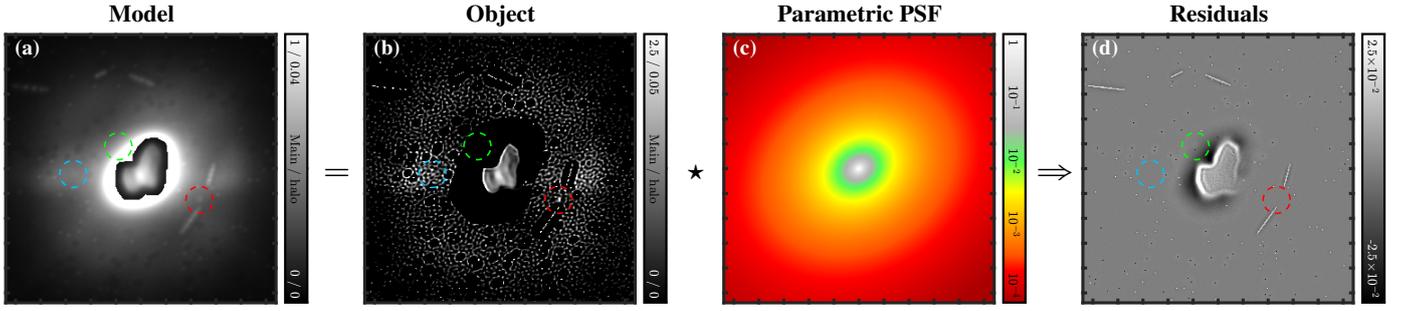
 
        \centering
        
        \newcommand{\PathFig}{Fig_overview/}
        \newcommand{\FigSuf}{_deconv}

        \newcommand{\LineRatio}{0.95}
        
        \newcommand{\fontTxt}[1]{\textbf{\small #1}}
        
        \newcommand{\widthTxt}{16pt}
        
        \newcommand{\sizeTxt}[1]{\Large{#1}}
        
        \newcommand{\spaceLine}{1.8cm}
        
        \newcommand{\widthFig}{\dimexpr (\linewidth - \widthTxt * 3)}
        
        \newcommand{\subfigColor}{white}

        \newcommand{\FigOne}{\PathFig Data\FigSuf}
        \newcommand{\FigTwo}{\PathFig Obj\FigSuf}
        \newcommand{\FigThree}{\PathFig PSF\FigSuf}
        \newcommand{\FigFour}{\PathFig Res\FigSuf}
        \newcommand{\FigOneBar}{\PathFig Data_bar}
        \newcommand{\FigTwoBar}{\PathFig Obj_bar}
        \newcommand{\FigThreeBar}{\PathFig PSF_bar}
        \newcommand{\FigFourBar}{\PathFig Res_core_bar}
        
        \sbox1{\includegraphics{\FigOne}}
        \sbox2{\includegraphics{\FigOneBar}}
        \sbox3{\includegraphics{\FigTwo}}
        \sbox4{\includegraphics{\FigTwoBar}}
        \sbox5{\includegraphics{\FigThree}}
        \sbox6{\includegraphics{\FigThreeBar}}
        \sbox7{\includegraphics{\FigFour}}
        \sbox8{\includegraphics{\FigFourBar}}
        
        \newcommand{\ColumnWidth}[1]
                {\dimexpr \LineRatio \widthFig * \AspectRatio{#1} / (\AspectRatio{1} + \AspectRatio{2} + \AspectRatio{3} + \AspectRatio{4} + \AspectRatio{5} + \AspectRatio{6} + \AspectRatio{7} + \AspectRatio{8}) \relax
                }
        \newcommand{\ColumnGap}{\hspace {\dimexpr \widthFig /13 - \LineRatio\widthFig /13 }}

        \begin{tabular}{
                @{\ColumnGap}
                M{\ColumnWidth{1}}
                @{\ColumnGap}
                M{\ColumnWidth{2}}
                @{\ColumnGap}
                M{\widthTxt}
                @{\ColumnGap}
                M{\ColumnWidth{3}}
                @{\ColumnGap}
                M{\ColumnWidth{4}}
                @{\ColumnGap}
                M{\widthTxt}
                @{\ColumnGap}
                M{\ColumnWidth{5}}
                @{\ColumnGap}
                M{\ColumnWidth{6}}
                @{\ColumnGap}
                M{\widthTxt}
                @{\ColumnGap}
                M{\ColumnWidth{7}}
                @{\ColumnGap}
                M{\ColumnWidth{8}}
                @{\ColumnGap}
                }
                
                \fontTxt{Model}&&
                &
                \fontTxt{Object}&&
                &
                \fontTxt{Parametric PSF}&&
                &
                \fontTxt{Residuals}
                \\                
                
                \subfigimg[width=\linewidth,pos=ul,font=\fontfig{\subfigColor}]{$\;$(a)}{0.0}{\FigOne} &
                \subfigimg[width=\linewidth,pos=ul,font=\fontfig{\subfigColor}]{}{0.0}{\FigOneBar} &
                \sizeTxt{$=$} &
                \subfigimg[width=\linewidth,pos=ul,font=\fontfig{\subfigColor}]{$\;$(b)}{0.0}{\FigTwo} &
                \subfigimg[width=\linewidth,pos=ul,font=\fontfig{\subfigColor}]{}{0.0}{\FigTwoBar} &
                \sizeTxt{$\conv$} &
                \subfigimg[width=\linewidth,pos=ul,font=\fontfig{\subfigColor}]{$\;$(c)}{0.0}{\FigThree} &
                \subfigimg[width=\linewidth,pos=ul,font=\fontfig{\subfigColor}]{}{0.0}{\FigThreeBar} &
                \sizeTxt{$\Rightarrow$} &
                \subfigimg[width=\linewidth,pos=ul,font=\fontfig{\subfigColor}]{$\;$(d)}{0.0}{\FigFour} &
                \subfigimg[width=\linewidth,pos=ul,font=\fontfig{\subfigColor}]{}{0.0}{\FigFourBar}
        \end{tabular}
             
        \caption{\label{fig:deconv_obj} Step~2 -- Object deconvolution. The estimated PSF core~(\refpan{c}=\refsubfig{fig:deconv_core}{c}) was used to deconvolve the object~(\refpan{b}). \refpan{a}: Model of the convolution. \refpan{d}: Residuals. Colour bars: see \reffig{fig:deconv_model}.
        }
\end{figure*}

As is shown in \reffig{fig:deconv_obj}, the objective of this step is to deconvolve the image of the object,~$\Vobj$, assuming that the PSF,~$\Vpsf$, is known, for example from the fit of its core done in the previous section \refsec{sec:core}. It is a classical deconvolution problem solved by minimising the following cost function~\citep[see also,][]{Berdeu:22_Elektra}:
\begin{equation} 
\label{eq:cost_obj}
	\cost{obj}{\Vdata,\Vpsf,\Vweight}{\Vobj} {}\triangleq{}
	\dat{wls}{\Vdata, \Vpsf\conv\Vobj, \Vweight}
	+ 
	\regpar^{\Tag{obj}}\reg{obj}{\Vobj}
	\,,
\end{equation}
under the physical constraint that~$\Vobj\geq\V{0}$. This hard constraint also limits oscillating artefacts close to sharp edges, commonly seen in deconvolution problems~\citep{Marchis:06_moon_detection, Fetick:20_param_marignal}. $\regpar^{\Tag{obj}}$ is a hyperparameter to balance the regularisation, $\reg{obj}{\Vobj}$, compared to the data fidelity term. $\reg{obj}{\Vobj}$ was chosen to favour smooth objects with sharp edges, by encouraging the sparsity of spatial gradients~\citep{Rudin:92_TV, Charbonnier:97_TV}
\begin{equation} 
	\label{eq:reg_obj}
	\begin{aligned}
		\reg{obj}{\V{\varphi}} {}\triangleq{} &
		\\
		& \hspace{-20pt}
		\sum_{\x}\Brack{\sqrt{\Paren{\brack{\nabla_{1}\varphi\Paren{\x}}^{2}+\brack{\nabla_{2}\varphi\Paren{\x}}^{2}+\brack{\epsilon^{\Tag{obj}}}^{2}}}-\epsilon^{\Tag{obj}}}
		\,,
	\end{aligned}
\end{equation}
where~$\nabla_{1}$ and $\nabla_{2}$ correspond to finite difference operators along the first and second spatial dimensions of the image. $\epsilon^{\Tag{obj}}>0$ is a threshold controlling the transition between a $\ell^1$-norm (edge-preserving) and a $\ell^2$-norm (smoothness). $\epsilon^{\Tag{obj}}$ also ensures that \refeq{eq:reg_obj} is differentiable at zero. Such an edge-preserving regularisation or equivalents are common in asteroid deconvolution~\citep{Mugnier:04_Mistral, Fetick:20_param_marignal, Berdeu:22_Elektra, Lau:23_prior_AOPSF, Yan:23_myopic_MCMC}. In practice, these hyperparameters, $\regpar^{\Tag{obj}}$ and $\epsilon^{\Tag{obj}}$, were manually tuned. $\regpar^{\Tag{obj}}$ mainly depends on the object's characteristics. Since all asteroids share similar properties, it is not expected that this parameter changes a lot for resolved asteroids. $\epsilon^{\Tag{obj}}$ varies with the awaited contrast on the asteroid surface: (i) for a well-resolved asteroid, values of a few percent of the object's dynamics provide good results, (ii) for poorly resolved targets where surface details are hardly expected, values of around a thousandth or less better favour sharp-edged binary reconstructions.

At this stage, there is still no constraint on potential moons. They are thus in the deconvolved model and disappear from the residuals, as is shown in \refsubfig{fig:deconv_obj}{d}. Once the object is deconvolved, it can be segmented with a threshold to remove the moons or other artefacts from the image model, as is described in \refapp{app:algo_deconv}, \refalg{alg:obj_wing}.

\subsection{Step 3. Deconvolution of the extended PSF wings}
\label{sec:wing}

\begin{figure*}[t!]
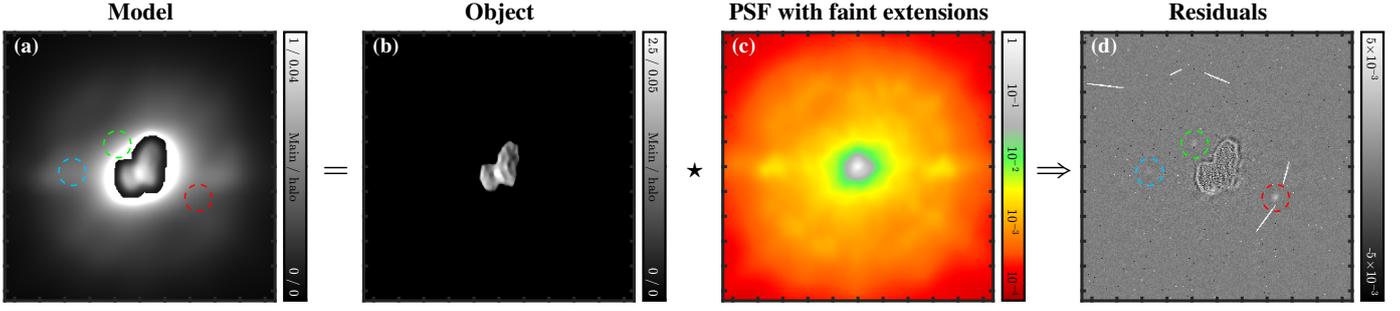
 
        \centering
        
        \newcommand{\PathFig}{Fig_overview/}
        \newcommand{\FigSuf}{_wing}

        \newcommand{\LineRatio}{0.95}
        
        \newcommand{\fontTxt}[1]{\textbf{\small #1}}
        
        \newcommand{\widthTxt}{16pt}
        
        \newcommand{\sizeTxt}[1]{\Large{#1}}
        
        \newcommand{\spaceLine}{1.8cm}
        
        \newcommand{\widthFig}{\dimexpr (\linewidth - \widthTxt * 3)}
        
        \newcommand{\subfigColor}{white}

        \newcommand{\FigOne}{\PathFig Data\FigSuf}
        \newcommand{\FigTwo}{\PathFig Obj\FigSuf}
        \newcommand{\FigThree}{\PathFig PSF\FigSuf}
        \newcommand{\FigFour}{\PathFig Res\FigSuf}
        \newcommand{\FigOneBar}{\PathFig Data_bar}
        \newcommand{\FigTwoBar}{\PathFig Obj_bar}
        \newcommand{\FigThreeBar}{\PathFig PSF_bar}
        \newcommand{\FigFourBar}{\PathFig Res_bar}
        
        \sbox1{\includegraphics{\FigOne}}
        \sbox2{\includegraphics{\FigOneBar}}
        \sbox3{\includegraphics{\FigTwo}}
        \sbox4{\includegraphics{\FigTwoBar}}
        \sbox5{\includegraphics{\FigThree}}
        \sbox6{\includegraphics{\FigThreeBar}}
        \sbox7{\includegraphics{\FigFour}}
        \sbox8{\includegraphics{\FigFourBar}}
        
        \newcommand{\ColumnWidth}[1]
                {\dimexpr \LineRatio \widthFig * \AspectRatio{#1} / (\AspectRatio{1} + \AspectRatio{2} + \AspectRatio{3} + \AspectRatio{4} + \AspectRatio{5} + \AspectRatio{6} + \AspectRatio{7} + \AspectRatio{8}) \relax
                }
        \newcommand{\ColumnGap}{\hspace {\dimexpr \widthFig /13 - \LineRatio\widthFig /13 }}

        \begin{tabular}{
                @{\ColumnGap}
                M{\ColumnWidth{1}}
                @{\ColumnGap}
                M{\ColumnWidth{2}}
                @{\ColumnGap}
                M{\widthTxt}
                @{\ColumnGap}
                M{\ColumnWidth{3}}
                @{\ColumnGap}
                M{\ColumnWidth{4}}
                @{\ColumnGap}
                M{\widthTxt}
                @{\ColumnGap}
                M{\ColumnWidth{5}}
                @{\ColumnGap}
                M{\ColumnWidth{6}}
                @{\ColumnGap}
                M{\widthTxt}
                @{\ColumnGap}
                M{\ColumnWidth{7}}
                @{\ColumnGap}
                M{\ColumnWidth{8}}
                @{\ColumnGap}
                }
                
                \fontTxt{Model}&&
                &
                \fontTxt{Object}&&
                &
                \fontTxt{PSF with faint extensions}&&
                &
                \fontTxt{Residuals}
                \\                
                
                \subfigimg[width=\linewidth,pos=ul,font=\fontfig{\subfigColor}]{$\;$(a)}{0.0}{\FigOne} &
                \subfigimg[width=\linewidth,pos=ul,font=\fontfig{\subfigColor}]{}{0.0}{\FigOneBar} &
                \sizeTxt{$=$} &
                \subfigimg[width=\linewidth,pos=ul,font=\fontfig{\subfigColor}]{$\;$(b)}{0.0}{\FigTwo} &
                \subfigimg[width=\linewidth,pos=ul,font=\fontfig{\subfigColor}]{}{0.0}{\FigTwoBar} &
                \sizeTxt{$\conv$} &
                \subfigimg[width=\linewidth,pos=ul,font=\fontfig{\subfigColor}]{$\;$(c)}{0.0}{\FigThree} &
                \subfigimg[width=\linewidth,pos=ul,font=\fontfig{\subfigColor}]{}{0.0}{\FigThreeBar} &
                \sizeTxt{$\Rightarrow$} &
                \subfigimg[width=\linewidth,pos=ul,font=\fontfig{\subfigColor}]{$\;$(d)}{0.0}{\FigFour} &
                \subfigimg[width=\linewidth,pos=ul,font=\fontfig{\subfigColor}]{}{0.0}{\FigFourBar}
        \end{tabular}
             
        \caption{\label{fig:deconv_wing} Step~3 -- PSF wing deconvolution. The segmented object~(\refpan{b}) was used to deconvolve the PSF wings~(\refpan{c}). \refpan{a}: Model of the convolution. \refpan{d}: Residuals. Colour bars: see \reffig{fig:deconv_model}.
        }
\end{figure*}

Now that a de-blurred image of the main object,~$\Vobj$, has been correctly estimated, it is possible to revert the problem and use this image as the convolution kernel to deconvolve the AO-PSF wings, as is shown in \reffig{fig:deconv_wing}. In this approach, the moons, which are a priori unknown, are not explainable by a model of an extended object convolved with a PSF, and should thus appear in the residuals; so should defective or hot pixels or random cosmic rays impacting the sensor, which cannot be calibrated in advance. These outliers should not corrupt the reconstruction of the PSF wings. As was introduced in \refsec{sec:intro}, the mean square error of \refeq{eq:WLS} is consequently not adapted. I propose to tackle this problem with robust penalisation~\citep[][Chap. 5]{Zoubir:18_robust_stat, Flasseur:19_PhD}. This consists of replacing the quadratic norm of the residuals in \refeq{eq:WLS} with a function,~$\rob$, that limits the influence of large (and aberrant) values in the minimisation:
\begin{equation} 
	\label{eq:robdata}
	\robdata{\V{\varphi}_{1},\V{\varphi}_{2},\V{w}}
	{}\triangleq{}
	\sum_{\x} \rob\Paren{\sqrt{w\Paren{\x}}\Paren{\varphi_{1}\Paren{\x} -\varphi_{2}\Paren{\x}}}
	\,.
\end{equation}

Of the different robust estimators,~$\rob$, available in the literature~\citep{Holland:77}, I chose the Cauchy function:
\begin{equation}
	\rob\Paren{r}
	{}\triangleq{}
	\frac{\gamma^2}{2}\ln\Paren{1+ r^2 / \gamma^2}\,.
\end{equation}
If the confidence model,~$\Vweight$ in \refeq{eq:robdata}, is properly scaled and calibrated so that the argument
\begin{equation}
r\Paren{\x} = \sqrt{w\Paren{\x}}\Paren{\varphi_{1}\Paren{\x} -\varphi_{2}\Paren{\x}}
\end{equation} follows a normalised Gaussian noise statistics (centred on zero and of a standard deviation equal to one) in the absence of outliers, setting~$\gamma = 2.385$ ensures that minimising \refeq{eq:WLS} or \refeq{eq:robdata} gives similar results. In doing so, the Cauchy function has the advantages of having no tuning parameter and of being differentiable.

In more detail, 
\begin{equation} 
	\label{eq:Cauchy_weight}
	\weightrob\Paren{r}=\frac{1}{r}\frac{\partial\rob\Paren{r}}{\partial r}=\Paren{1+r^2 / \gamma^2}^{-1}
	\,,
\end{equation}
the so-called robust weight in the following, is the correction factor to apply to the weights in the least square error of \refeq{eq:WLS} to make it equivalent to the robust penalisation of \refeq{eq:robdata}. This is typically used in the context of IRLS, as was introduced in \refsec{sec:intro}:
\begin{equation} 
	\label{eq:weight_IRLS}
	\weight^\Tag{irls}\Paren{\x} = \weightrob\Paren{r\Paren{\x}}\weight\Paren{\Vx}
	\,.
\end{equation}

\refsubfigfull{fig:weight}{a} gives an example of a robust weight map, obtained by applying $\weightrob$ of \refeq{eq:Cauchy_weight} to a weighted halo residual map when fitting the full PSF,~$\Vpsf$. The outliers can directly be identified in black, with equivalent weights tending towards zero. It is then possible to discard them on the fly, using a conservative threshold,~$\throb$ (typically \percent{50}, as is discussed in \refapp{app:algo_deconv}), and as is shown in \refsubfig{fig:weight}{b}. This is done by updating \refeq{eq:weight}, with the additional remark that this equation is in fact only an approximation. Indeed, the noisy data,~$\Vdata$, is not the real flux. A better estimate of the weighting confidence term,~$\Vweight$, can be obtained by replacing~$\Vdata$ with the forward model
\begin{equation}
	\label{eq:data_mod}
	\Vdata^{\Tag{mod}} = \Vpsf \conv\Vobj
\end{equation}
in \refeq{eq:weight}. In total, as is detailed in \refalg{alg:obj_wing} of \refapp{app:algo_deconv}, the updated weight map now reads as
\begin{equation}
	\label{eq:weight_mod}
	\weight\Paren{\Vx} =
	\begin{cases}
		0 \text{ if } \weightrob\Paren{\Vx} \leq \throb
		\\
		1/\Paren{\eta\data^{\Tag{mod}}\Paren{\Vx}+\varRON} \text{ otherwise}
	\end{cases}
	\,.
\end{equation}
It is iteratively updated in an approach similar to IRLS, \refeq{eq:weight_IRLS}, but with (i) a binary update when it comes to the robust estimator, and (ii) a refinement of the expected noise level with the intensity predicted by the forward model. Doing so prevents any further corruption of the deconvolved PSF wings by strong outliers.

\begin{figure}[t!] 
        \centering
        
        \newcommand{\PathFig}{Fig_weight/}
        
        \newcommand{\LineRatio}{0.925}
        
        \newcommand{\fontTxt}[1]{\textbf{\small #1}}
        
        \newcommand{\subfigColor}{black}        
        
        \newcommand{\FigOne}{\PathFig Weight_map}  
        \newcommand{\FigTwo}{\PathFig Valid_map}  
        \newcommand{\FigThree}{\PathFig Weight_bar}  
        \sbox1{\includegraphics{\FigOne}}
        \sbox2{\includegraphics{\FigTwo}}
        \sbox3{\includegraphics{\FigThree}}
        
        \newcommand{\ColumnWidth}[1]
                {\dimexpr \LineRatio \linewidth * \AspectRatio{#1} / (\AspectRatio{1} + \AspectRatio{2} + \AspectRatio{3}) \relax
                }
        \newcommand{\ColumnGap}{\hspace {\dimexpr \linewidth /4 - \LineRatio\linewidth /4 }}

        \begin{tabular}{
                @{\ColumnGap}
                M{\ColumnWidth{1}}
                @{\ColumnGap}
                M{\ColumnWidth{2}}
                @{\ColumnGap}
                M{\ColumnWidth{3}}
                @{\ColumnGap}
                }
                
                \fontTxt{Robust weights}&
                \fontTxt{Discarded pixels}&
                \\                
                
                \subfigimg[width=\linewidth,pos=ul,font=\fontfig{\subfigColor}]{$\;$(a)}{0.0}{\FigOne} &
                \subfigimg[width=\linewidth,pos=ul,font=\fontfig{\subfigColor}]{$\;$(b)}{0.0}{\FigTwo} &
                \subfigimg[width=\linewidth,pos=ul,font=\fontfig{\subfigColor}]{}{0.0}{\FigThree}
                \\

        \end{tabular}        
        \caption{\label{fig:weight} Automatic removal of outliers with respect to the model. \refpan{a}: Equivalent weights of the robust penalisation,~$\weightrob\paren{\sqrt{\Vweight}\paren{\Vdata - \Vdata^{\Tag{mod}}}}$. \refpan{b}: Discarding of pixels (black) below a threshold,~$\throb$.}
\end{figure}

The cost function that is iteratively minimised to reconstruct the PSF, $\Vpsf$, is
\begin{equation} 
	\label{eq:cost_wing}
	\cost{psf}{\Vdata, \Vobj, \Vweight}{\Vpsf}
	{}\triangleq{}
	\dat{wls}{\Vdata,\Vpsf\conv\Vobj,\Vweight}
	+ \regpar^{\Tag{psf}}\reg{psf}{\Vpsf}
	\,,
\end{equation}
under the physical constraint that~$\Vpsf \geq 0$. $\regpar^{\Tag{psf}}$ is a hyperparameter to balance $\reg{psf}{\Vpsf}$. I define this regularisation term as
\begin{equation} 
	\label{eq:reg_psf}
	\reg{psf}{\V{\varphi}}
	{}\triangleq{}
	\sum_{\x}\Brack{\nabla_{1}\Brack{\ln\Paren{\varphi\Paren{\x}}}}^{2}+\Brack{\nabla_{2}\Brack{\ln\Paren{\varphi\Paren{\x}}}}^{2}
	\,,
\end{equation}
which consists of the classical $\ell^2$-norm on the gradient but applied to the logarithm of the PSF. The former ensures a smooth reconstruction, while the latter acts as a `dynamic whitening' term. Indeed, the challenge in regularising the PSF wings is that $\Vpsf$ spans multiple orders of magnitude between the PSF core vicinity and the external turbulence halo, as is seen in \refsubfig{fig:deconv_model}{c}. $\reg{psf}{\Vpsf}$ must act through this high dynamic range. Similar to $\Vweight$ in \refeqs{eq:WLS}{eq:robdata}, which scales the residuals according to the admissible awaited noise level (noise whitening), the regularisation must scale the PSF wings. Since
\begin{equation}
	\nabla_{i}\Brack{\ln\Paren{\varphi\Paren{\x}}} = \Brack{\nabla_{i}\varphi\Paren{\x}} / \varphi\Paren{\x}
	\,,
\end{equation}
I let the opportunity for the gradients in \refeq{eq:reg_psf} to evolve proportionally to the local intensity of the total PSF, despite the large dynamics of the unknowns, both refining the PSF core and recovering its wings. As for $\regpar^{\Tag{obj}}$ in \refsec{sec:deconv}, the hyperparameter $\regpar^{\Tag{psf}}$ is a function of the noise level of the data,~$\Vdata$, and was manually tuned in this study.

Technically, one could stop here if the main objective were to retrieve the faint moon hidden in the bright halo. Nonetheless, now that a better model of the AO-PSF is known, it is possible to loop back on the object deconvolution step and iterate the steps of \refsecs{sec:deconv}{sec:wing} to further refine the reconstruction while robustly updating the noise model,~$\Vweight$, via \refeq{eq:weight_mod} in \refeqs{eq:cost_obj}{eq:cost_wing}. This implementation is further detailed in \refalg{alg:obj_wing} of \refapp{app:algo_deconv}. This proved to be efficient for complex PSF cores, badly described by a 2D Moffat pattern, for example in the presence of motion blur or for bad seeing conditions. On top of a better object deconvolution, this also improves the residual quality close to the object edges, potentially unveiling moons very close to the main body.

\section{Validation on simulations}
\label{sec:res_sim}

The method was tested on simulated data representative of real conditions. The simulated object, shown in \refsubfig{fig:deconv_model}{b}, is an image of 67P/Churyumov–Gerasimenko obtained by Rosetta (European Space Agency), downscaled to the resolution and size (between 100 and 200 milliarcseconds) that are typical of the targets observed within the ESO Large Programme described in \ref{sec:res_data}. As was mentioned in \refsec{sec:method}, the PSF used to blur the object image is a real AO-PSF obtained with ZIMPOL on a star. Three moons with different contrasts and distances were added to the main body (coloured circles in \refsubfig{fig:deconv_model}{d}). The photon shot noise and the readout noise parameters, $\V{\eta}$ and $\VvarRON$, are assumed to be identical for all the pixels in the FoV:
\begin{equation}
	\label{eq:weight_const}
	\forall \Vx \in \text{FOV}, \eta\Paren{\Vx} = \eta \text{ and } \varRON\Paren{\Vx} = \varRON
	\,.
\end{equation}
On top of these noises, random pixels are modelled as hot or dead (`salt \& pepper' noise) and five cosmic rays are randomly simulated (orange arrows).

\begin{figure}[t!] 
        \centering
        
        \newcommand{\PathFig}{Fig_sim_analysis/}
        
        \newcommand{\spaceLine}{-0.05cm}
        
        \newcommand{\LineRatio}{0.95}
        \newcommand{\subfigColor}{white}        
        
        \newcommand{\FigOne}{\PathFig Main_true}  
        \newcommand{\FigTwo}{\PathFig Main_BM}  
        \newcommand{\FigThree}{\PathFig Main_core} 
        \newcommand{\FigFour}{\PathFig Main_wing} 
        \sbox1{\includegraphics{\FigOne}}
        \sbox2{\includegraphics{\FigTwo}}
        \sbox3{\includegraphics{\FigThree}}
        \sbox4{\includegraphics{\FigFour}}
        
        \newcommand{\ColumnWidth}[1]
                {\dimexpr \LineRatio \linewidth * \AspectRatio{#1} / (\AspectRatio{1} + \AspectRatio{2} + \AspectRatio{3} + \AspectRatio{4}) \relax
                }
        \newcommand{\ColumnGap}{\hspace {\dimexpr \linewidth /5 - \LineRatio\linewidth /5}}
                
        \begin{tabular}{
                @{\ColumnGap}
                M{\ColumnWidth{1}}
                @{\ColumnGap}
                M{\ColumnWidth{2}}
                @{\ColumnGap}
                M{\ColumnWidth{3}}
                @{\ColumnGap}
                M{\ColumnWidth{4}}
                @{\ColumnGap}
                }         
                
                \subfigimg[width=\linewidth,pos=ul,font=\fontfig{\subfigColor}]{$\,$(a)}{0.0}{\FigOne}
                &
                \subfigimg[width=\linewidth,pos=ul,font=\fontfig{\subfigColor}]{$\,$(b)}{0.0}{\FigTwo}
                &
                \subfigimg[width=\linewidth,pos=ul,font=\fontfig{\subfigColor}]{$\,$(c)}{0.0}{\FigThree}
                &
                \subfigimg[width=\linewidth,pos=ul,font=\fontfig{\subfigColor}]{$\,$(d)}{0.0}{\FigFour}
                \\[\spaceLine]
                \subfigimg[width=\linewidth,pos=ul,font=\fontfig{\subfigColor}]{$\,$(e)}{0.0}{\PathFig Main_blurred}
                &
                \subfigimg[width=\linewidth,pos=ul,font=\fontfig{\subfigColor}]{$\,$(f)}{0.0}{\PathFig Res_BM}
                &
                \subfigimg[width=\linewidth,pos=ul,font=\fontfig{\subfigColor}]{$\,$(g)}{0.0}{\PathFig Res_core}
                &
                \subfigimg[width=\linewidth,pos=ul,font=\fontfig{\subfigColor}]{$\,$(h)}{0.0}{\PathFig Res_wing}
                \\[\spaceLine]
                \subfigimg[width=\linewidth,pos=ul,font=\fontfig{\subfigColor}]{$\,$(i)$\;\;\,\kappa=\percent{100.2}$}{0.0}{\PathFig Res_deconv_obj}
                &
                \subfigimg[width=\linewidth,pos=ul,font=\fontfig{\subfigColor}]{$\,$(j)$\;\;\,\kappa=\percent{114.1}$}{0.0}{\PathFig Res_deconv_BM}
                &
                \subfigimg[width=\linewidth,pos=ul,font=\fontfig{\subfigColor}]{$\,$(k)$\;\;\kappa=\percent{106.7}$}{0.0}{\PathFig Res_deconv_core}
                &
                \subfigimg[width=\linewidth,pos=ul,font=\fontfig{\subfigColor}]{$\,$(l)$\quad\;\kappa=\percent{97.3}$}{0.0}{\PathFig Res_deconv_wing}
                \\[\spaceLine]
        \end{tabular}
        
        \renewcommand{\LineRatio}{0.9}
        \renewcommand{\subfigColor}{black}        
        
        \renewcommand{\FigOne}{\PathFig Profiles_main}  
        \renewcommand{\FigTwo}{\PathFig Profiles_PSF}  
        \sbox1{\includegraphics{\FigOne}}
        \sbox2{\includegraphics{\FigTwo}}
        
        \renewcommand{\ColumnWidth}[1]
                {\dimexpr \LineRatio \linewidth * \AspectRatio{#1} / (\AspectRatio{1} + \AspectRatio{2}) \relax
                }
        \renewcommand{\ColumnGap}{\hspace {\dimexpr \linewidth /3 - \LineRatio\linewidth /3 }}
                
        \begin{tabular}{
                @{\ColumnGap}
                M{\ColumnWidth{1}}
                @{\ColumnGap}
                M{\ColumnWidth{2}}
                @{\ColumnGap}
                }         
                
                \subfigimg[width=\linewidth,pos=ul,font=\fontfig{\subfigColor}]{$\quad\!$(m)}{0.0}{\FigOne}
                &
                \subfigimg[width=\linewidth,pos=ul,font=\fontfig{\subfigColor}]{$\quad\;$(n)}{0.0}{\FigTwo}
                \\
        \end{tabular}

        \caption{\label{fig:sim_analysis} Comparison of the deconvolution with the simulated truth. \refpan{a}:~Simulated true object,~$\Vobj$. \refpan{b}:~Estimated binary object,~$\bar{\Vdata}$, applying a threshold to \refpan{e}. \refpan{c}:~The object's first deconvolution after the estimation of the PSF core parameters,~$\pcore$. \refpan{d}:~The object's final deconvolution after the alternate estimation of the PSF wing,~$\Vpsf$. \refpan{e}:~Simulated blurred and noisy data,~$\Vdata$. \refpans{f-h}:~Model residuals,~$\Vdata - \Vdata^{\Tag{mod}}$, for the binary object (\refpan{f}), after the first deconvolution (\refpan{g}) and after the final deconvolution (\refpan{h}). \refpans{i-l}:~Thrice the absolute value of the object residuals with the shifted object,~$3\Abs{\tilde{\Vobj} - \kappa\Vobj^{\Tag{mod}}}$, for the true object (\refpan{i}), for the binary object (\refpan{j}), after the first deconvolution (\refpan{k}), and after the final deconvolution (\refpan{l}). Intensity scale of \refpans{a-e,i-l}: see \refsubfig{fig:deconv_model}{b}. Intensity scale of \refpans{f-h}: see \refsubfig{fig:deconv_core}{d}. \refpan{m}:~Intensity slices along the dashed coloured line of \refpans{a-e}. \refpan{n}:~Radial profiles of the true PSF (black), its closest Moffat fit (red), the PSF core estimate (blue), and the PSF wing deconvolution (green) on a logarithmic scale.}
\end{figure}

The results of the different steps of the method are shown in Figs.~\ref{fig:deconv_core},~\ref{fig:deconv_obj}, and~\ref{fig:deconv_wing}. \reffigfull{fig:sim_analysis} focuses on the main body deconvolution, providing zooms on the deconvolved images,~$\Vobj^{\Tag{mod}}$, the model residuals,~$\Vdata - \Vdata^{\Tag{mod}}$, and the object residuals,~$\tilde{\Vobj} - \kappa\Vobj^{\Tag{mod}}$. $\tilde{\Vobj}$ stands for the shifted true object,~$\Vobj$. Indeed, due to the PSF core deformation, the PSF in \refsubfig{fig:deconv_model}{c} may not be properly centred. By construction (see \refapp{app:algo}), my method reconstructs a centred PSF. As a consequence, the deconvolved object may be translated compared to the truth. This shift is estimated by fitting a perfect 2D Moffat model in the true PSF. $\tilde{\Vobj}$ is the cubic interpolation of~$\Vobj$ translated by the position of this fitted 2D Moffat pattern. \refsubfigfull{fig:sim_analysis}{i} shows that this shift is vertical, producing the bright arcs at the top and the bottom of the object residuals. To emphasise the surface texture retrieval rather than a bias in the global intensity, the deconvolved~$\Vobj^{\Tag{mod}}$ is further normalised with
\begin{equation}
	\kappa = \argmin{\tilde{\kappa}}{\sum_{\Vx} \Abs{\obj^{\Tag{mod}}\Paren{\Vx}-\tilde{\kappa}\tilde{\obj}\Paren{\Vx}}}
	\,,
\end{equation}
using the absolute value in the sum to be robust to outlier regions, badly deconvolved or smoothed by the deconvolution such as edges or very bright or dark areas.

\reffigfull{fig:deconv_core} shows that the oversimplified model of the first step efficiently fits the bright inner part of the halo that is correctly removed from the residuals of \refsubfig{fig:deconv_core}{d} and \refsubfig{fig:sim_analysis}{f}. By design, the outer part of the structured halo, mainly driven by the AO cut-off frequency, is still present in the residuals, as it cannot be explained by this simple model. As for the object, the threshold applied to the data produces a binary object (see \refsubfig{fig:deconv_core}{b} and \refsubfig{fig:sim_analysis}{b}, whose edges are consistent with the blurred object, as is shown by the orange contour in \refsubfig{fig:deconv_core}{a}). This is further confirmed by the slices of \refsubfig{fig:sim_analysis}{m}, the edges of the binary approximation (dark grey) being within a few pixels of the true slice of the object (light grey). As was expected, the overall object intensity is underestimated by more than~\percent{10} because the energy diluted by the PSF wings is not integrated and of course none of the surface structure is retrieved, \refsubfig{fig:sim_analysis}{j}. As for the PSF core, the radially averaged profile of the blindly estimated Moffat (blue curve in \refsubfig{fig:sim_analysis}{n}) shows a very good agreement with the theoretical radial profile of the 2D Moffat (red curve) fitted directly in the true PSF.

\reffigfull{fig:deconv_obj} shows that using the fitted PSF core is indeed sufficient to retrieve the morphology of the object, as is discussed in \refsec{sec:core}. Structures at the surface of the deconvolved object, \refsubfig{fig:deconv_obj}{b} and \refsubfig{fig:sim_analysis}{c}, are consistent with the simulated truth, \refsubfig{fig:deconv_model}{b} and \refsubfig{fig:sim_analysis}{a}, as is shown by the good residuals of \refsubfig{fig:sim_analysis}{k}. Nonetheless, the fact that the Moffat model cannot properly grasp the deformations of the PSF core produces two kinds of artefacts: (i)~a bright edge on the left side of the object in \refsubfigs{fig:sim_analysis}{c,k} that suggests that this edge is `over-deconvolved', and (ii)~non-homogeneous structures in the residuals at the edge's location, shown in \refsubfig{fig:sim_analysis}{g}. This can also be seen in the blue slice in \refsubfig{fig:sim_analysis}{m}, which shows unnatural deconvolution peaks at the edges of the reconstructed object. And as was already discussed, the fact that the PSF core does not account for the PSF wings induces an underestimation of the object's intensity, here by more than ~\percent{6}. I also emphasise that, by construction, the halo is found in this deconvolved image of \refsubfig{fig:deconv_obj}{b}, as well as the moons and other noise artefacts. They consequently appear in the modelled data, \refsubfig{fig:deconv_obj}{a}, and the moons are not visible in the residuals, \refsubfig{fig:deconv_obj}{d}.

All of these problems are solved after a few iterations of the alternate algorithm, once the PSF wings are properly deconvolved, as is shown by \reffig{fig:deconv_wing}. Concerning the object, the zooms of \refsubfigs{fig:sim_analysis}{d,l} and the green slice of \refsubfig{fig:sim_analysis}{m} show that, despite being slightly smoothed by the regularisation of \refeq{eq:cost_obj}, the details are quantitatively retrieved with the correct intensity within~\percent{$100.2-97.3 \leq 3$}. The sharp edges are also well defined at the correct location (see the green and light grey curves). Compared with \refsubfigs{fig:sim_analysis}{c,k}, the `over-deconvolution' problem on the left side of the object is greatly improved. Nonetheless, the combination of (i)~the threshold to segment the main object, (ii)~the regularisation in \refeq{eq:cost_obj}, and (iii)~the approximate knowledge of the PSF core (or equivalently the cut-off of the optical transfer function of the system) within a blind approach, limits the precision of the object's edge recovery, as is visible in the data residuals of \refsubfig{fig:sim_analysis}{h}. In addition, the spatial regularisation erases the finest details on the surface of the object, as is seen in the model residuals of \refsubfig{fig:sim_analysis}{l}. These are known limits of the edge-preserving regularisation~\citep{Fetick:20_param_marignal} but are a necessary trade-off with the PSF reconstruction to control the alternate blind deconvolution convergence. Otherwise, the residuals reach the noise level, below the percent of the data dynamic (see colour bars of \refsubfig{fig:deconv_wing}{d} for \refsubfig{fig:sim_analysis}{h}). There is no noticeable feature at the location of the object's surface that would imply an over-regularisation in the deconvolution, indicating that the details smoothed by the regularisation are indeed lost in the noise level.

Concerning the PSF, all the features mentioned in \refsec{sec:method} are quantitatively retrieved in \refsubfig{fig:deconv_wing}{c}. The wind-driven halo is well defined. The brightest structures on the AO cut-off frequency ring and the diffraction spikes are clearly visible and deconvolved. Only the resolution on the speckle pattern is smoothed by the regularisation but its global structure has the correct dynamics. In \refsubfig{fig:sim_analysis}{n}, the radial profile of the deconvolved PSF perfectly matches the radial profile obtained on the true PSF. As is explained in \refsec{sec:wing}, the moons and other outliers are robustly penalised in the equivalent weights in \refsubfig{fig:weight}{a} and the brightest ones are successfully rejected from the fit, \refsubfig{fig:weight}{b}. Thus, they do not impact the model of the halo, \refsubfig{fig:deconv_wing}{a}, and are consequently nicely visible in the residuals, \refsubfig{fig:deconv_wing}{d}. Compared with the injected signals in \refsubfig{fig:deconv_model}{d}, the moons are slightly dimmed and erased from the data, showing that the method suffers from slight self-subtraction. This is nonetheless a cost worth paying to efficiently retrieve the halo and constitutes a trade-off in the regularisation of \refeq{eq:cost_wing} between the reconstruction of the PSF structures and the moon rejection from their deconvolution.

\section{Examples in real data}
\label{sec:res_real}

\subsection{The multiple asteroids \Kleopatra and \Elektra}
\label{sec:res_data}

The ESO Large Programme `Asteroids as tracers of Solar System formation: Probing the interior of primordial main belt asteroids'~\citep[ID 199.C-0074, PI:][R filter, $\lambda=\unit{645.9}{\nano\meter}$, $\Delta\lambda=\unit{56.7}{\nano\meter}$]{Vernazza:21_Large_program} targeted several asteroids of the Solar System's main belt in order to characterise the albedo, surface, shape, and mass of the ones accompanied by moons. The data were obtained with ZIMPOL, mounted on the Spectro Polarimetric High-contrast Exoplanet REsearcher~\citep[SPHERE,][]{Beuzit:19_SPHERE} of the Very  Large Telescope (VLT) observatory, equipped with the SPHERE Adaptive eXtreme Optics system~\citep[SAXO,][]{Fusco:16_SAXO}. Mainly designed for polarimetry acquisitions, ZIMPOL has an imaging mode. Working in visible wavelengths, it offers an unprecedented resolution of small Solar System bodies at the cost of a more complex AO-PSF. I applied my method to two targets of this Large Programme: \Kleopatra and \Elektra. 

Among the \unit{100}{\kilo\meter} class asteroids, \Kleopatra is quite original. It has a dumbbell shape and is orbited by two known moons, which constrain its gravitational field, and thus its mass and density \citep{Ostro:00_Kleopatra, Descamps:11_Kleopatra_triple, Shepard:18_Kleopatra_radar, Marchis:21_Kleopatra}. Its elongated shape and the availability of state-of-the-art deconvolutions and 3D reconstructions make it a good case with which to test the deconvolution performance of the proposed method.

\Elektra is also an asteroid of the main belt, surrounded by three moons~\citep{Fuksa:23_Elektra_orbit}. Its faintest and closest companion was missed until recently due to the lack of proper data reduction and data analysis tools~\citep{Yang:16_Elektra_Minerva, Berdeu:22_Elektra}. It was detected in archival data obtained in 2014 with the Integral Field Spectrograph of SPHERE~\citep[IFS,][]{Claudi:08_SPHERE_IFS}. The complexity of ZIMPOL AO-PSFs and the variability of the brightness and distance of the different moons relative to the main body represent a challenging case with which to test the proposed method. \Elektra was observed at different epochs in 2018 and 2019 by \cite{Vernazza:21_Large_program}, each epoch gathering multiple frames. In order to assess the consistency of the method results, both in terms of deconvolution and moon enhancement, two different frames of four epochs were selected. To show the versatility of the method and the large acceptability of its assumptions, I also selected additional datasets from other instruments: the IFS (ID 60.A-9362(A), PI: Yang et al. 2014, YJH filter, $\lambda$ from 0.95 to \unit{1.65}{\micro\meter}), the InfraRed Dual-band Imager and Spectrograph \citep[IRDIS,][ID 296.C-5038(A), PI: Yang et al. 2017, DK12 filter, $\lambda=\unit{2.11}{\micro\meter}$, $\Delta\lambda=\unit{102}{\nano\meter}$]{Dohlen:08_IRDIS} of SPHERE, and the Near Infrared Camera 2 (NIRC2, ID U58N2, PI: de~Pater et al. 2005, Kp filter, $\lambda=\unit{1.248}{\micro\meter}$, $\Delta\lambda=\unit{163}{\nano\meter}$) of the Keck II telescope.

\subsection{Main body deconvolution and surface detail recovery}

\begin{figure}[t!] 
        \centering
        
        \newcommand{\PathFig}{Fig_Elektra_profiles/}
        
        \newcommand{\LineRatio}{0.95}
        
        \newcommand{\fontTxt}[1]{\textbf{\small #1}}
        
        \newcommand{\widthTxt}{12pt}
        
        \newcommand{\sizeTxt}[1]{\Large{#1}}
        
        \newcommand{\spaceLine}{-1pt}
        
        \newcommand{\widthFig}{\dimexpr (\linewidth - \widthTxt)}
        
        \newcommand{\subfigColor}{white}        
        
        \sbox1{\includegraphics{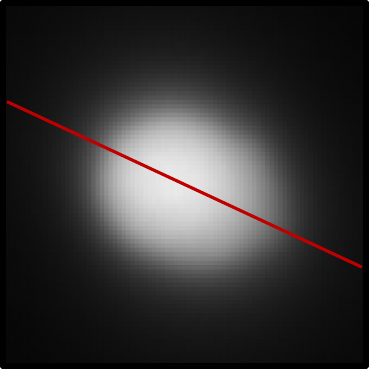}}
        \sbox2{\includegraphics{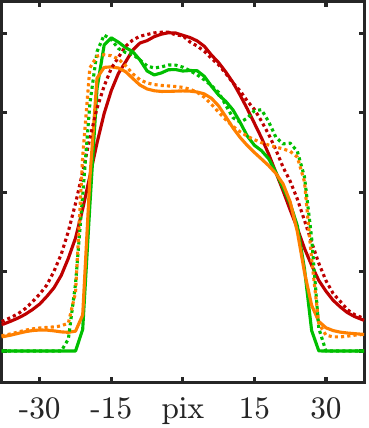}}
        
        \newcommand{\ColumnWidth}[1]
                {\dimexpr \LineRatio \widthFig * \AspectRatio{#1} / (\AspectRatio{1}*3/2 + \AspectRatio{2}) \relax
                }
        \newcommand{\ColumnGap}{\hspace {\dimexpr \widthFig /6 - \LineRatio\widthFig /6 }}
        
        \newcommand{\InsertCol}[3]{
            \begin{tabular}{
                @{}
    		C{\linewidth}
                @{}
    		}
    		\subfigimg[width=0.99\linewidth,pos=ul,font=\fontfig{\subfigColor}]{#2}{0.0}{\PathFig #1_1}
\\[-2pt]	\subfigimg[width=0.99\linewidth,pos=ul,font=\fontfig{\subfigColor}]{#3}{0.0}{\PathFig #1_2}
    	\end{tabular}
        }        
        
		\newcommand{\InsertLine}[1]{
			\rotatebox[origin=l]{90}{\fontTxt{#1}}
			&
			\InsertCol{#1_Main_Data}{(1)}{(2)}
			&
			\InsertCol{#1_Main_LAM}{}{}
			&
			\InsertCol{#1_Main_wing}{}{}
			&
			\subfigimg[width=\linewidth,pos=ul,font=\fontfig{\subfigColor}]{}{0.0}{\PathFig #1_Profiles}
			\\[\spaceLine]
        }

        \begin{tabular}{
                @{\ColumnGap}
                M{\widthTxt}
                @{\ColumnGap}
                B{\ColumnWidth{1}/2}
                @{\ColumnGap}
                B{\ColumnWidth{1}/2}
                @{\ColumnGap}
                B{\ColumnWidth{1}/2}
                @{\ColumnGap}
                M{\ColumnWidth{2}}
                @{\ColumnGap}
                }
                
            &
            \fontTxt{Data}
            &
            \fontTxt{Reference}
            &
            \fontTxt{Method}
            &
            \fontTxt{Slices}
			\\[\spaceLine]
                
 			\InsertLine{2019-07-30}
 			\InsertLine{2019-08-04}
			\InsertLine{2019-08-05}
 			\InsertLine{2019-08-06}       
 			         
        \end{tabular}        
        \caption{\label{fig:Elektra_profiles} Zoom on the deconvolved image of \Elektra in \reffig{fig:comp_instru} obtained with ZIMPOL. For information, the deconvolutions (green) are compared with the data (red) and the reference (orange) obtained~\cite{Vernazza:21_Large_program}. To emphasise the gain in resolution, all the figures are normalised to the same linear intensity scale. For each epoch, two different frames (plain (1) and dashed (2) curves) are given.}
\end{figure}

\reffigsfull{fig:Elektra_profiles}{fig:Kleopatra} gather the results obtained on \Elektra and \Kleopatra for different epochs with different seeing conditions. The noise model parameters,~$\paren{\eta, \varRON}$, of \refeq{eq:weight}, critical to scale the robust penalisation, are empirically adjusted in the data, as is described in \refapp{app:noise_model}.

\begin{figure*}[t!] 
        \centering
        
        \newcommand{\PathFig}{Fig_Kleopatra/}
        
        \newcommand{\LineRatio}{0.95}
        
        \newcommand{\fontTxt}[1]{\textbf{\small #1}}
        
        \newcommand{\widthTxt}{12pt}
        
        \newcommand{\sizeTxt}[1]{\Large{#1}}
        
        \newcommand{\spaceLine}{-1pt}
        
        \newcommand{\widthFig}{\dimexpr (\linewidth - \widthTxt)}
        
        \newcommand{\subfigColor}{white}        
        
        \sbox1{\includegraphics{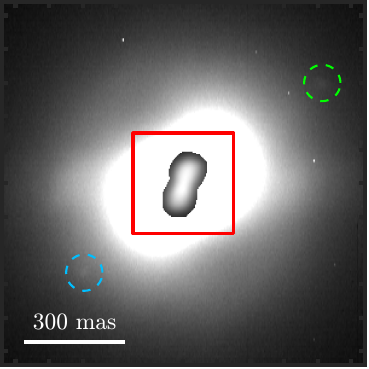}}
        \sbox2{\includegraphics{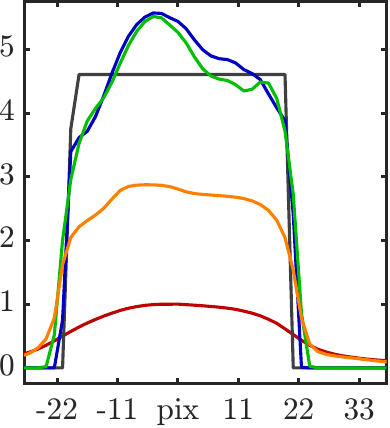}}
        
        \newcommand{\ColumnWidth}[1]
                {\dimexpr \LineRatio \widthFig * \AspectRatio{#1} / (\AspectRatio{1}*10/2 + \AspectRatio{2}) \relax
                }
        \newcommand{\ColumnGap}{\hspace {\dimexpr \widthFig /10 - \LineRatio\widthFig /10 }}
        
        \newcommand{\InsertCol}[4]{
            \begin{tabular}{
                @{}
    		C{\linewidth}
                @{}
    		}
    		\subfigimg[width=0.975\linewidth,pos=ul,font=\fontfig{\subfigColor}]{#3}{0.0}{\PathFig #1}
\\[-1pt]	\subfigimg[width=0.975\linewidth,pos=ul,font=\fontfig{\subfigColor}]{#4}{0.0}{\PathFig #2}
    	\end{tabular}
        }        
        
		\newcommand{\InsertLine}[9]{
			\rotatebox[origin=l]{90}{\fontTxt{#1}}
			&
			\InsertCol{#1_Main_Data}{#1_Main_LAM}{#2}{#3}
			&
			\InsertCol{#1_Main_BM}{#1_Res_BM}{#4}{#5}
			&
			\InsertCol{#1_Main_core}{#1_Res_core}{#6}{#7}
			&
			\InsertCol{#1_Main_wing}{#1_Res_wing}{#8}{#9}
			&
			\subfigimg[width=\linewidth,pos=ul,font=\fontfig{\subfigColor}]{}{0.0}{\PathFig #1_Profiles}
			&
			\subfigimg[width=\linewidth,pos=ul,font=\fontfig{\subfigColor}]{}{0.0}{\PathFig #1_Data}
			&
			\subfigimg[width=\linewidth,pos=ul,font=\fontfig{\subfigColor}]{}{0.0}{\PathFig #1_PSF}
			&
			\subfigimg[width=\linewidth,pos=ul,font=\fontfig{\subfigColor}]{}{0.0}{\PathFig #1_Res}
			\\[\spaceLine]
        }

        \begin{tabular}{
                @{\ColumnGap}
                M{\widthTxt}
                @{\ColumnGap}
                B{\ColumnWidth{1}/2}
                @{\ColumnGap}
                B{\ColumnWidth{1}/2}
                @{\ColumnGap}
                B{\ColumnWidth{1}/2}
                @{\ColumnGap}
                B{\ColumnWidth{1}/2}
                @{\ColumnGap}
                M{\ColumnWidth{2}}
                @{\ColumnGap}
                M{\ColumnWidth{1}}
                @{\ColumnGap}
                M{\ColumnWidth{1}}
                @{\ColumnGap}
                M{\ColumnWidth{1}}
                @{\ColumnGap}
                }
                
            &
            \multicolumn{4}{c}{\fontTxt{Data / deconvolution / residuals}}
            &
            \fontTxt{Slices}
            &
            \fontTxt{Data}
            &
            \fontTxt{Deconvolved PSF}
            &
            \fontTxt{Halo residuals}
			\\[\spaceLine]
                
 			\InsertLine{2017-07-14}{(a)}{(b)}{(c)}{(d)}{(e)}{(f)}{(g)}{(h)}
			\InsertLine{2017-07-27}{}{}{}{}{}{}{}{}
			\InsertLine{2017-08-10}{}{}{}{}{}{}{}{}
			\InsertLine{2017-08-22}{}{}{}{}{}{}{}{}
 			\InsertLine{2018-12-22}{}{}{}{}{}{}{}{}
 			         
        \end{tabular}        
        \caption{\label{fig:Kleopatra} Examples of deconvolution on \Kleopatra at different epochs. \textit{First column}~--~Zoom on the main body. \refpans(a,b):~Data (top, red slice) and reference deconvolution (bottom, orange slice) obtained by \cite{Vernazza:21_Large_program}. \refpans(c,d):~Estimated binary object after the estimation of the PSF core parameters (top, grey slice) and residuals (bottom). \refpans(e,f):~Deconvolution with the PSF core (top, blue slice) and residuals (bottom). \refpans(g,h):~Deconvolution after the alternate estimation of the PSF wing (top, green slice) and residuals (bottom). To highlight the faint intensities, all the object representations share the same square root stretch intensity scale normalised to the data's maximal value. The residuals are displayed with the colour scale of \refsubfig{fig:deconv_core}{b}. \textit{Second column}~--~Slices along the coloured lines of the first column. \textit{Third column}~--~Data displayed with the dual linear scale of \refsubfig{fig:deconv_model}{a}. \textit{Fourth column}~--~PSF displayed with the colour scale of \refsubfig{fig:deconv_model}{c} (peak normalised to one). \textit{Fifth column}~--~Residuals of the halo model displayed with the colour scale of \refsubfig{fig:deconv_model}{d}. For information, the deconvolved main body image was inserted.}
\end{figure*}

\reffigfull{fig:Elektra_profiles} focuses on the final deconvolutions of \Elektra, obtained after the alternate algorithm. The reconstructed edges are sharp without any `corona artefact' (see below discussion on \reffig{fig:Kleopatra}) or motion blur residual, even for 2019-07-30 (2), which is a challenging case, as is highlighted by the reference deconvolution or the important deformation of the PSF core in \reffig{fig:comp_instru}. Nonetheless, for the 2019-08-05 frames, the overshoot on the left side of the asteroid also indicates a slight over-deconvolution, as was already seen in the simulations, \refsec{sec:res_sim}. Features at the asteroid's surface are resolved, as is seen in the zooms of \reffig{fig:Elektra_profiles}. The evolution of the projected shadows on the surface is visible between the frames of the best epochs, 2019-08-05 and 2019-08-06, and the corresponding slices of \reffig{fig:Elektra_profiles}. This supports the facts that the method is consistent and that these are not deconvolution artefacts. For 2019-08-06, the slices show that the contrast in the structures decreases between the two frames, as the land forms face the Sun more. Except for the main shadow of 2019-08-05, none of these details can be seen in the reference deconvolution obtained by \cite{Vernazza:21_Large_program}.

\reffigfull{fig:Kleopatra} details the different steps of the blind deconvolution algorithm on \Kleopatra. For each epoch, the first columns show a zoom of the data, the fitted binary model, and the deconvolved images based on the knowledge of the PSF core or the PSF wings (and the corresponding residuals) as well as intensity slices along the coloured lines. The deconvolutions obtained by \cite{Vernazza:21_Large_program} are also given to serve as references, \refpans{b}. \refpans{c} show that the binary threshold of the data provides a good approximation of the object support, both in terms of size and shape. Then, it clearly appears that the mere knowledge of the PSF core is sufficient to retrieve the object body, but the deconvolved images of \refpans{f} suffer from what is sometimes called `corona artefact', a stepwise artefact surrounding the object edges that originates from a coupling between the edge-preserving regularisation and an incorrect description of the PSF core~\citep{Marchis:06_moon_detection, Fetick:19_Vesta, Fetick:20_param_marignal, Lau:23_prior_AOPSF}. These images are similar to the reference reconstructions. For some epochs, such as 2017-07-27, 2017-08-10, and 2017-08-22, the artefacts surrounding the object strongly suggest hiccups of the AO loop or even motion blur. This could be explained by a shift in the telescope pointing direction in between the four Detector Integration Times (DITs), which are averaged in one ZIMPOL-reduced frame by the data reduction pipeline~\citep{Schmid:18_ZIMPOL}. This is particularly visible in the blue slices of 2017-07-27 and 2017-08-10 that cut along this motion blur.

Deconvolving the full PSF solves these issues. In the final step of the proposed method, the edges of \Kleopatra are de-blurred and sharp, and the duplicated images are collapsed into a single image in the cases of strong motion blur. This is also emphasised by the green slices in \reffig{fig:Kleopatra}. These deconvolved images obtained via a blind deconvolution algorithm with limited priors on the object and the PSF compare well with state-of-the-art marginal deconvolutions and are consistent with the previously derived 3D model of the asteroid~\citep{Shepard:18_Kleopatra_radar, Marchis:21_Kleopatra, Vernazza:21_Large_program, Lau:23_prior_AOPSF}. The orange slices of \reffig{fig:Kleopatra} show that the reference method underestimates the object's intensity and can only retrieve smooth and large-scale structures. Due to these limitations, only a difference in albedo between the two lobes of the asteroid was studied by~\cite{Marchis:21_Kleopatra}. The small-scale structures in the green slices of \reffig{fig:Kleopatra} suggest that finer details from \Kleopatra{}'s surface can be retrieved with my method.

Interestingly, the deconvolution residuals in \reffig{fig:Kleopatra} show a line-by-line horizontal stripe pattern. This can be directly linked to the ZIMPOL imaging mode where only one out of two lines of the sensor is exposed. To produce `high-resolution' images, the reduction pipeline interpolates the missing lines~\citep{Schmid:18_ZIMPOL}. This implies that the proposed deconvolution algorithm reaches the level of the artefacts introduced by the data reduction. To limit the impact of this `zebra' in the deconvolved image, I had to increase the regularisation parameter, $\regpar^{\Tag{obj}}$, compared to \refsec{sec:res_sim}, limiting the performance of the method of edge and detail recovery. This explains why spatial features can still be seen in the residuals. These reduction features that correlate neighbouring pixels also mean that the assumption of independent noise previously mentioned in \refsec{sec:core} for \refeq{eq:weight_diag} is not strictly correct and that the diagonal weighting term, $\Vweight$, in \refeq{eq:cost_wing} should actually be the inverse of a covariance matrix, as in \refeq{eq:WLS_full}. This effect was nonetheless neglected in this study.

\subsection{Point spread function reconstruction and halo removal}

A quick look at the halo brightness in the data column of \reffig{fig:Kleopatra} shows that the turbulence conditions change between the different epochs. This directly impacts the deconvolved AO-PSF wings and their contrast. For the two epochs 2017-07-14 and 2017-08-10, the PSF is even mainly dominated by the wind-driven halo. Nonetheless, the blind PSF deconvolution successfully retrieves the AO cut-off limit at the correct location as well as its brightest speckles, despite the absence of an instrumental prior in the PSF model. For good seeing conditions, in spite of the elongated shape of \Kleopatra, the reconstructed speckles appear nicely resolved and split. In addition, the orientation and elongation of the wind-driven halo do not appear to correlate with \Kleopatra's orientation (see especially 2017-07-14 and 2017-08-22), as one could have feared from a potential cross-talk during the blind deconvolution. This overall consistency favours the robustness of the method in connection with the shape of the object and with the turbulence conditions.

\begin{figure*}[p!] 
        \centering
        
        \newcommand{\PathFig}{Fig_comp_instru/}
        
        \newcommand{\LineRatio}{0.95}
        
        \newcommand{\fontTxt}[1]{\textbf{\small #1}}
        
        \newcommand{\widthTxt}{12pt}
        
        \newcommand{\sizeTxt}[1]{\Large{#1}}
        
        \newcommand{\spaceLine}{-2pt}
        
        \newcommand{\widthFig}{\dimexpr (\linewidth - \widthTxt)}
        
        \newcommand{\subfigColor}{white}        
        
        \sbox1{\includegraphics{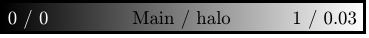}}
        
        \newcommand{\ColumnWidth}
                {\dimexpr \LineRatio \widthFig * \AspectRatio{1} / (\AspectRatio{1}*5) \relax
                }
        \newcommand{\ColumnGap}{\hspace {\dimexpr \widthFig /7 - \LineRatio\widthFig /7 }}    
        
		\newcommand{\InsertLine}[3]{
			\rotatebox[origin=l]{90}{\fontTxt{#2}}
			&
			\subfigimg[width=\linewidth,pos=ul,font=\fontfig{\subfigColor}]{#3}{0.0}{#1_Data}
			&
			\subfigimg[width=\linewidth,pos=ul,font=\fontfig{\subfigColor}]{}{0.0}{#1_Model}
			&
			\subfigimg[width=\linewidth,pos=ul,font=\fontfig{\subfigColor}]{}{0.0}{#1_PSF}
			&
			\subfigimg[width=\linewidth,pos=ul,font=\fontfig{\subfigColor}]{}{0.0}{#1_Res}
			&
			\subfigimg[width=\linewidth,pos=ul,font=\fontfig{\subfigColor}]{}{0.0}{#1_Weight}
			\\[\spaceLine]
        }

        \begin{tabular}{
                @{\ColumnGap}
                M{\widthTxt}
                @{\ColumnGap}
                M{\ColumnWidth}
                @{\ColumnGap}
                M{\ColumnWidth}
                @{\ColumnGap}
                M{\ColumnWidth}
                @{\ColumnGap}
                M{\ColumnWidth}
                @{\ColumnGap}
                M{\ColumnWidth}
                @{\ColumnGap}
                }
                
            &
            \fontTxt{Data}
            &
            \fontTxt{Halo and object models}
            &
            \fontTxt{Point spread function}
            &
            \fontTxt{Halo residuals}
            &
            \fontTxt{Robust weights}
			\\
                
			\InsertLine{\PathFig/ZIMPOL_2019-07-30_2}{ZIMPOL}{$\;$2019-07-30 (2)}
			\InsertLine{\PathFig/ZIMPOL_2019-08-04_2}{ZIMPOL}{$\;$2019-08-04 (2)}
			\InsertLine{\PathFig/ZIMPOL_2019-08-05_2}{ZIMPOL}{$\;$2019-08-05 (2)}
			\InsertLine{\PathFig/ZIMPOL_2019-08-06_1}{ZIMPOL}{$\;$2019-08-06 (1)}
			\InsertLine{\PathFig/IFS}{IFS}{$\;$2014-12-30}
			\InsertLine{\PathFig/IRDIS}{IRDIS}{$\;$2016-02-16}
			\InsertLine{\PathFig/NIRC2}{NIRC2}{$\;$2005-01-15}
			\InsertLine{\PathFig/Bar}{}{}
 			         
        \end{tabular}

        \begin{tabular}{
                @{\ColumnGap}
                M{\widthTxt}
                @{\ColumnGap}
                M{\ColumnWidth}
                @{\ColumnGap}
                M{\widthTxt}
                @{\ColumnGap}
                M{\ColumnWidth}
                @{\ColumnGap}
                M{\widthTxt}
                @{\ColumnGap}
                M{\ColumnWidth}
                @{\ColumnGap}
                M{\widthTxt}
                @{\ColumnGap}
                M{\ColumnWidth}
                @{\ColumnGap}
                }
                
            \rotatebox[origin=l]{90}{\fontTxt{PSF on a star}}
			&
            \subfigimg[width=\linewidth,pos=ul,font=\fontfig{\subfigColor}]{$\;$ZIMPOL}{0.0}{\PathFig/ZIMPOL_2019-07-30_1_PSF_star}
            &
            \rotatebox[origin=l]{90}{}
            &
            \subfigimg[width=\linewidth,pos=ul,font=\fontfig{\subfigColor}]{$\;$IFS}{0.0}{\PathFig/IFS_PSF_star}
            &
            \rotatebox[origin=l]{90}{}
            &
            \subfigimg[width=\linewidth,pos=ul,font=\fontfig{\subfigColor}]{$\;$IRDIS}{0.0}{\PathFig/IRDIS_PSF_star}
            &
            \rotatebox[origin=l]{90}{}
            &
            \subfigimg[width=\linewidth,pos=ul,font=\fontfig{black}]{$\;$NIRC2}{0.0}{\PathFig/NIRC2_PSF_star}
			\\[\spaceLine]
 			         
        \end{tabular}

        \caption{\label{fig:comp_instru} Examples of robust halo removal on \Elektra with different instruments (see also \reffig{fig:Elektra_data}). \textit{First (resp. second) column}~--~Blurred data, $\Vdata$ (resp. deconvolved image, $\Vobj$, and halo model, $\Vdata^{\Tag{mod}} = \Vpsf\conv\Vobj$), normalised and displayed with a dual linear scale. \textit{Third column}~--~Deconvolved PSF, $\Vpsf$ (peak normalised to one). \textit{Fourth column}~--~Residuals of the halo model, $\Vdata - \Vdata^{\Tag{mod}}$. \textit{Fifth column}~--~Map of the equivalent robust weights, $\weightrob\Paren{\sqrt{\Vweight}\Paren{\Vdata - \Vdata^{\Tag{mod}}}}$. Orange arrows: Camera defects, dead pixel clusters, or cosmic rays. When visible, the three moons of \Elektra are circled. The red arrows point towards the outer moon when it is outside the FoV. \textit{Last line}~--~Examples of PSF on a star for each instrument.}
\end{figure*}

Looking at \reffigs{fig:comp_instru}{fig:Elektra_data}, the conclusions are similar with \Elektra, with the additional supporting fact that for each epoch the two reconstructed PSFs are extremely consistent in terms of turbulence features (contrast and wind-driven halo). Furthermore, the rotation of the spider diffraction spikes is clearly visible (see 2019-07-30 and 2019-08-06 in particular).

After halo removal, the two moons of \Kleopatra are clearly visible at all epochs in the halo residuals, as is shown in the last column of \reffig{fig:Kleopatra} (coloured circles). The transition between a readout noise-limited regime to a photon shot noise-limited regime is also nicely visible: the noise level increases in the object's vicinity where the halo is very bright in the data, especially for bad seeing conditions (2017-07-14, 2017-08-10, and 2017-08-22). Looking at the halo residuals in \reffig{fig:comp_instru}, this transition is also visible, especially for the bad seeing cases of 2019-07-30 and 2019-08-04 with an increased noise level in the object's vicinity. The three moons of \Elektra are visible in all the presented ZIMPOL frames (coloured circles), even when the deconvolution is not perfect (see 2019-07-30). Indeed, by construction, even if the object and the PSF are not perfectly split, moons (point sources) remain outliers of the model and should appear in the residuals. This is an encouraging result that shows the robustness of the approach. The counterparts of the moons in the robust weight maps are also nicely visible. The orange arrows point towards discarded outliers, such as defective pixels or random cosmic rays as well as camera defects impacting some columns of the detector.

The last lines of \reffig{fig:comp_instru} present results on data obtained with other instruments and reference PSFs measured on a star. They show that the method is robust for a large variety of AO-PSFs.

Looking at the IFS line, my method is able to reproduce the results of my previous pragmatic approach~\citep{Berdeu:22_Elektra}, extracting the three moons from the diffuse halo. The poor AO performance of this dataset explains why the diffraction rings of the reconstructed PSF differ from the IFS reference. Nonetheless, the diffraction spikes are visible and the reconstructed multi-lobed core nicely matches the PSF produced by the brightest moon (red circles): it presents a strong secondary lobe, at the top left of the main core. A similar complex PSF core is retrieved with ZIMPOL for the 2019-07-30 (2) frame (secondary lobe on the right). This shows that the method is, to some extent, robust for multi-lobed PSF cores diverging from a simple Moffat pattern, slightly relaxing this potential limitation of the method. The robust weights obtained with the IFS appear quite low. This comes from the wrong noise model fitted in the reduced data via the method presented in \refapp{app:noise_model}. Indeed, the reduction performed by PIC of IFS data~\citep{Berdeu:20_PIC} is based on a spatial regularisation that biases and correlates the noise in the reconstruction. Such an incorrect noise model can be handled by relaxing the constraint on $\throb$ to lower values for automatic outlier rejection.

The IRDIS data are very noisy, with a dense pollution with defective pixels (common with infrared detectors), while \Elektra only spans a few pixels. Despite this unfavourable situation, the first Airy rings of the PSF are reconstructed with the correct dynamics and with negligible signal in the residuals. The first ring is fragmented due to poor seeing conditions. Unfortunately, only the brightest moon is visible because of the high noise level and high density of unusable pixels.

Finally, with NIRC2, \Elektra is barely resolved, with an angular size similar to that of the first diffraction ring of the instrument PSF. Despite this challenging situation, the characteristic hexagonal shape of this ring, due to the fragmentation of the Keck~II mirror with hexagonal segments, is nicely recovered. So are the secondary diffraction peaks (two in green at the top right of the core for the first order and a few in yellow for the second order) that produce replicated images of \Elektra in the saturated data. Concerning the moons, only the brightest is visible in the residuals, highlighting the need for more frames and epochs and dedicated detection algorithms to increase the signal-to-noise ratio.

\section{Conclusions and perspectives}
\label{sec:conclu}

In this paper, I have presented a novel blind deconvolution technique that can retrieve both the object and the faint wings of AO-PSFs. The only main priors on the object are that it be contrasted with a smooth shape and sharp edges. Contrary to most of the existing blind deconvolution algorithms, it is not based on a parametric model of the PSF, allowing one to reconstruct a large variety of shapes and structures, as featured by PSFs after an AO correction. The proposed inverse problem approach is an alternate deconvolution with an edge-preserving regularisation for the object and a regularisation on the smoothness of the gradients for the PSF with intensity whitening. Tested on simulated and real data, my method outperforms state-of-the-art techniques \citep{Fetick:19_model_based_AOPSF, Vernazza:21_Large_program, Lau:23_prior_AOPSF, Yan:23_myopic_MCMC}, by both efficiently reconstructing details at the surface of the objects and faithfully retrieving the features of the AO-PSFs. It produces a physical and realistic model of the bright halo surrounding the object, which can then be carefully removed. This enhances faint moons in its close vicinity with a better contrast and at closer distances than previous halo-subtraction methods \citep{Assafin:08_digital_coronography, Yang:16_Elektra_Minerva, Pajuelo:18_Carry_coronagraph, Berdeu:22_Elektra}. 

Implemented in a general framework and with a limited number of assumptions, my method can be applied to different instruments mounted on different telescopes equipped with different AO systems. It paves the way for the study of resolved asteroids or moon systems around the giant planets of our Solar System via archival data of different instruments for which previous methods could not reveal the companions. It also prepares the arrival of the future generation of ELTs, whose complex and sensitive PSFs and AO systems will require robust and versatile processing techniques. Nonetheless, as is discussed below, there are several ways to further improve this method.

Solving the dual deconvolution problem implies carefully tuning the method. This is mainly done through the hyperparameters of the regularisations on the object and the PSF, namely $\regpar^{\Tag{obj}}$, $\epsilon^{\Tag{obj}}$, $\regpar^{\Tag{psf}}$, and $\throb$. In particular, one could decide to first keep a strong regularisation on the object to prioritise the PSF reconstruction, and to relax it the second time, once the PSF has been correctly estimated, to better retrieve the object's details. Providing an automatic tuning of such parameters to get a fully unsupervised algorithm is an active field of research~\citep{The:22_SURE} and implementing such methods was beyond the scope of this paper. In this work, these parameters were manually tuned and fixed once and for all for each instrument. The fact that they worked for all the tested datasets is encouraging and shows their good admissible range in terms of noise level, object shape, and seeing conditions. This thus limits the need for further manual tuning by the user.

It was emphasised in \refsec{sec:res_real} that lots of datasets suffer from poor AO-PSF quality or even motion blur. Robustness is obtained via a higher regularisation level, but this can degrade the image quality or produce over-deconvolution. To increase the resolution and the deconvolution quality, and since the targets are very bright, one should work on the individual ZIMPOL DITs rather than the stacked image. The results also show that the method reaches the limits of the ZIMPOL reduction pipeline, whose artefacts dominate the residuals in some datasets. Implementing such changes implies modifying the reduction pipeline and was beyond the scope of this paper.

As is shown in \refsec{sec:res_real}, most of the outliers, such as defective pixels or cosmic rays, are successfully and robustly identified and discarded on the fly by the proposed algorithm. Nonetheless, some of them are missed, and so are the faintest moons. As is discussed in \refsec{sec:res_sim}, the proposed method consequently suffers from slight self-subtraction, which could dim the signal of potential moons. Even if it is less pronounced than with other methods \citep{Yang:16_Elektra_Minerva, Pajuelo:18_Carry_coronagraph}, some precautions must be taken when fitting the photometry of a moon that can be underestimated, or when announcing the absence of moons below a given noise-limited contrast. It is possible to assess these biases by injecting false moons as is done in the standard exoplanet detection algorithm to quantitatively estimate the contrast performance~\citep{Flasseur:18_PACO}. This was beyond the scope of the paper, but should be kept in mind for further studies.

Finally, the natural next step of this method is to use to the reconstructed PSF within matching filter algorithms to detect even fainter moons in the residuals. Combined with the refined noise model and with the robust weights to limit false positive detections by discarding outliers, this would push the detection limits even further. Echoing the challenges of exoplanet detection, it comes nonetheless with the additional difficulty of the rapid motion of the moons around the main object~\citep{Showalter:19_Neptune, Berdeu:22_Elektra} and is an open research field.



\begin{acknowledgements}
	The author would like to warmly thank the Referee Laurent Mugnier for his pertinent feedback and numerous remarks and suggestions that greatly contributed to improving the quality and clarity of the article.
	\\
	The author would like to thank Éric Thiébaut and Ferréol Soulez for the fruitful discussions on robust detection algorithms and adequate prior for high dynamic unknowns as well as Benoît Carry for his careful proofreading.
	\\
	All the ZIMPOL reductions are based on public data acquired for the ESO Large Programme ID 199.C-0074 (Vernazza et al. 2021, `Asteroids as tracers of Solar System formation: Probing the interior of primordial main belt asteroids') and available at \href{https://observations.lam.fr/astero/}{https://observations.lam.fr/astero/
}.
	\\
	The IFS and IRDIS reductions are based on public data provided by the ESO Science Archive Facility. The author would like to thank Antoine Kaszczyc and Maud Langlois who provided the PSFs and reduced the IRDIS raw data.
	\\
	The NIRC2 reductions are based data provided by the Keck Observatory Archive (KOA), which is operated by the W. M. Keck Observatory and the NASA Exoplanet Science Institute (NExScI), under contract with the National Aeronautics and Space Administration. The author would like to thank Kate Minker who reduced the raw data and provided the PSF.
	\\
	This project has received funding from the European Union's Horizon 2020 research and innovation programme under grant agreement No 101004719.
\end{acknowledgements}


\bibliographystyle{aa}
\bibliography{bib_AO_PSF_deconv}

\flushcolsend


\begin{appendix}

\section{Pseudo-codes of the method}
\label{app:algo}

This appendix gathers the pseudo-codes of the different steps of the proposed method, described in \refsec{sec:method}. In practice, all the code is implemented in MATLAB within the open-source \mbox{GlobalBioIm} framework~\citep{code:GlobalBioIm_19}. I also detail in this appendix how the different cost functions are minimised.

\subsection{Estimation of the point spread function core}
\label{app:algo_core}

The different parameters of the first step of \refsec{sec:core} were alternately fitted, as was described in \refalg{alg:core}. Indeed, applying a threshold to the data is not a differentiable operation, complicating the minimisation with iterative gradient descent methods. A first set of parameters of the PSF core was guessed, \reflin{core}{first_guess}, on a coarse grid of threshold values, \reflin{core}{rough_set}. These values were then refined by alternating five fits on (i) the threshold on the data, along with the amplitude parameter, \reflin{core}{obj_thr}, and (ii) the PSF core parameters, \reflin{core}{PSF_fit}. The optimisation problems were solved with the simplex search method of \cite{Lagarias:98_fminsearch}. For each problem, 200 iterations were performed. The parameters of the Moffat profiles were initialised with $\alpha = 3$ and $\beta=1.6$, which are standard values~\citep[see][]{Fetick:19_Vesta}. To reweight,~$\Vweight$, \reflin{core}{reweight}, a general threshold of \percent{2.5} was applied to the data. From the perspective of the dynamic of the PSF, this value roughly corresponds to the transition between its wind-driven core and its wings (see \refsubfig{fig:deconv_model}{c}). Only the meaningful pixels in the brightest part of the halo, not shaded in red,  \refsubfig{fig:deconv_core}{a}, are thus kept in the data fidelity term.

\begin{algorithm}
\caption{\label{alg:core} Fit of the object threshold and PSF core.}
\begin{algorithmic}[1]
\small

	\State $\weight\Paren{\Vx}
		\underseteq{eq:weight}{\gets} 
		\begin{cases}
			0 \text{ if } \Vdata\Paren{\Vx} \leq \percent{2.5}\f{max}{\data}
			\\
			1/\Paren{\eta\data\Paren{\Vx}+\varRON} \text{ otherwise}
        \end{cases}
        $
        \commentalgo{Confidence map}
		\label{lin:core:reweight}
        
	\State $\mathcal{D} \gets \Brace{\percent{30}, \percent{35}, \cdots, \percent{65}, \percent{70}}$
        \commentalgo{Set for rough initialisation}
		\label{lin:core:rough_set}
        
	\State $\Paren{\bar{d}, \gamma, \Vx_{0}, \V{\alpha}, \beta, \theta} \gets \argmin{d\in\mathcal{D}, \tilde{\gamma}, \tilde{\Vx}_{0}, \tilde{\V{\alpha}}, \tilde{\beta}, \tilde{\theta}}{\cost{core}{\Vdata, \V{w}}{d, \tilde{\gamma}, \tilde{\Vx}_{0}, \tilde{\V{\alpha}}, \tilde{\beta}, \tilde{\theta}}}$
        \commentalgo{$1^\Tag{st}$ guess}
		\label{lin:core:first_guess}
        
	\For{$i\text{ from } 1 \text{ to } 5$}
		\commentalgo{Alternate fit to minimise \refeq{eq:cost_core}}

		\State $\Paren{\bar{d}, \gamma} \gets \argmin{\tilde{d}, \tilde{\gamma}}{\cost{core}{\Vdata, \V{w}}{\tilde{d}, \tilde{\gamma}, \Vx_{0}, \V{\alpha}, \beta, \theta}}$
			\commentalgo{Object threshold}
			\label{lin:core:obj_thr}
			
		\State $\Paren{\gamma, \Vx_{0}, \V{\alpha}, \beta, \theta} \gets \argmin{\tilde{\gamma}, \tilde{\Vx}_{0}, \tilde{\V{\alpha}}, \tilde{\beta}, \tilde{\theta}}{\cost{core}{\Vdata, \V{w}}{\bar{d}, \tilde{\gamma}, \tilde{\Vx}_{0}, \tilde{\V{\alpha}}, \tilde{\beta}, \tilde{\theta}}}$
			\commentalgo{PSF core}
			\label{lin:core:PSF_fit}
			
	\EndFor
	\State \Return $\Paren{\bar{d}, \gamma, \Vx_{0}, \V{\alpha}, \beta, \theta}$
\end{algorithmic}
\end{algorithm}

\subsection{Alternate object and point spread function deconvolution}
\label{app:algo_deconv}

\refalg{alg:obj_wing} details the implementation of the dual deconvolution algorithm of the object and the PSF wings of \refsecs{sec:deconv}{sec:wing}. The role of \reflin{obj_wing}{th_obj} is to define the main object support. I empirically chose a threshold of $\bar{d}^\Tag{sup}$ between \percent{15} and \percent{25} of the maximal value of the median-filtered deconvolved object,~$\f{med}{\Vobj}$ (kernel of $5\times5$ pixels, as in \refapp{alg:noise}). Such a threshold is low enough to account for partially illuminated pixels on the object's edge, while being conservative enough to correctly constrain and tackle the `corona artefacts' mentioned in \refsec{sec:res_real}. In practice, this threshold was applied only in the vicinity of the main body. To further  avoid crenellated edges on the segmented image, this support was dilated by one pixel. The rest of the FoV was automatically excluded to avoid bright moons or artefacts, such as cosmic rays in the deconvolved image, exceeding this threshold.

Conversely, for \reflin{obj_wing}{w_rob_obj} and \reflin{obj_wing}{w_rob_wing}, the $\throb$ threshold was applied only in the halo to exclude the moons or artefacts, such as cosmic rays, but not in the main body vicinity, to keep this area (see \refsubfig{fig:weight}{b}). Indeed, this area is critical to fit the PSF core features but hard to fit due to the artificial threshold on the deconvolved image and the edge-preserving regularisation, as is seen in the robust weights of \reffig{fig:comp_instru}. I typically used the conservative value of \percent{50}, as it is important to remove any moon suspicion while the deconvolution of the smoothly varying PSF is robust for a high rate of missing data. In the presence of strong outliers (dead pixels) on the object support or its vicinity, as in \refsubfig{fig:weight}{b}, it is still possible to apply a less conservative threshold to this region, typically below \percent{10}.

The two minimisation problems of \reflin{obj_wing}{th_obj} and \reflin{obj_wing}{wing} were solved with 1000~iterations of the variable metric with limited memory-bounded algorithm \citep[VMLM-B,][]{Thiebaut:02}, a limited-memory quasi-Newton method with Broyden-Fletcher-Goldfarb-Shanno updates \citep[BFGS,][]{Nocedal:80} that handles bound constraints. In the case of strong motion blur, typically $\n{alt}=30$ alternate iterations are looped.

In \reflin{obj_wing}{w_rob_obj} and \reflin{obj_wing}{w_rob_wing}, the weights were robustly updated only after a few alternate loops,  $\n{wgt}=5$. I would like to remark here that if a proper calibration of the reduced data has been made, the robust weights, \reflin{obj_wing}{w_rob_0}, can be initialised with a map of already known aberrant pixels. This was done for the IRDIS dataset.

\begin{algorithm}
\caption{\label{alg:obj_wing} Alternate object and PSF wing deconvolution.}
\begin{algorithmic}[1]
\small

	\State $\Vpsf \gets \pcore\Paren{\V{0}, \V{\alpha}, \beta, \theta, \gamma}$
        \commentalgo{Initialisation of the PSF with the fitted core}
        
	\State $\Vweight^{\Tag{rob}} \gets \V{1}$
        \commentalgo{Initialisation of the robust weights without outliers}
		\label{lin:obj_wing:w_rob_0}
		
	\State $\weight\Paren{\Vx}
		\underseteq{eq:weight}{\gets} 1/\Paren{\eta\data\Paren{\Vx}+\varRON}$
		\commentalgo{Data-based confidence map}
        
	\For{$i\text{ from } 1 \text{ to } \n{alt}$}
		\commentalgo{$\n{alt}$ alternate fits to minimise \refeqs{eq:cost_obj}{eq:cost_wing}}
		
		\vspace{5pt}
		
		---- \textit{Object deconvolution} ----
		
		\vspace{2.5pt}	
		
		\If{$i>\n{wgt}$}
			\State $\weight\Paren{\Vx} \underseteq{eq:weight_mod}{\gets} 0 \text{ if } \weight^{\Tag{rob}}\Paren{\Vx} \leq \throb$
        			\commentalgo{Removing outliers}
				\label{lin:obj_wing:w_rob_obj}
		\EndIf
	
		\State $\Vobj
			\underseteq{eq:cost_obj}{\gets} \argmin{\tilde{\Vobj}\geq\V{0}}{\cost{obj}{\Vdata, \Vpsf, \Vweight}{\tilde{\Vobj}}}$
			\commentalgo{Object deconvolution}
	
		\State $\bar{\obj}\Paren{\Vx} \gets 
	        \begin{cases}
				0 \text{ if } \obj\Paren{\Vx} \leq \bar{d}^\Tag{sup}\f{max}{\f{med}{\Vobj}}
				\\
				\obj\Paren{\Vx} \text{ otherwise}
    	   	 \end{cases}
	        $
        	\hspace{-12pt}\commentalgo{Keeping the main body}
        	\label{lin:obj_wing:th_obj}
        	
		\State $\Vdata^{\Tag{mod}} \underseteq{eq:data_mod}{\gets} \Vpsf\conv\bar{\Vobj}$
        		\commentalgo{Updating model of the data}	
        		
		\State $\weight\Paren{\Vx}
					\underseteq{eq:weight_mod}{\gets} 1/\Paren{\eta\data^{\Tag{mod}}\Paren{\Vx}+\varRON}$
	       		\commentalgo{Model-based confidence map}
        		
        	\State $\Vweight^{\Tag{rob}}
			\underseteq{eq:Cauchy_weight}{\gets} \weightrob\Paren{\sqrt{\Vweight}\Paren{\Vdata - \Vdata^{\Tag{mod}}}}$
        	\commentalgo{Updating robust weights}

		\vspace{5pt}
		
		---- \textit{PSF deconvolution} ----
		
		\vspace{2.5pt}	
		
		\If{$i>\n{wgt}$}
			\State $\weight\Paren{\Vx} \underseteq{eq:weight_mod}{\gets} 0 \text{ if } \weight^{\Tag{rob}}\Paren{\Vx} \leq \throb$
        			\commentalgo{Removing outliers}
				\label{lin:obj_wing:w_rob_wing}
		\EndIf
		
		\State $\Vpsf
			\underseteq{eq:cost_wing}{\gets} \argmin{\tilde{\Vpsf}\geq \V{0}}{\cost{psf}{\Vdata, \bar{\Vobj},\Vweight}{\tilde{\Vpsf}}}$
			\commentalgo{PSF wing deconvolution}
			\label{lin:obj_wing:wing}

		\State $\Vdata^{\Tag{mod}} \underseteq{eq:data_mod}{\gets} \Vpsf\conv\bar{\Vobj}$
        		\commentalgo{Updating model of the data}

		\State $\weight\Paren{\Vx}
			\underseteq{eq:weight_mod}{\gets} 1/\Paren{\eta\data^{\Tag{mod}}\Paren{\Vx}+\varRON}$
			\commentalgo{Model-based confidence map}
		\State $\Vweight^{\Tag{rob}}
			\underseteq{eq:Cauchy_weight}{\gets} \weightrob\Paren{\sqrt{\Vweight}\Paren{\Vdata - \Vdata^{\Tag{mod}}}}$
        	\commentalgo{Updating robust weights}
        	
	\EndFor
	\State \Return $\Paren{\Vobj, \Vpsf, \Vdata^{\Tag{mod}}, \Vweight^{\Tag{rob}}}$
\end{algorithmic}
\end{algorithm}

\section{Empirical noise model}
\label{app:noise_model}

\begin{figure}[t!]
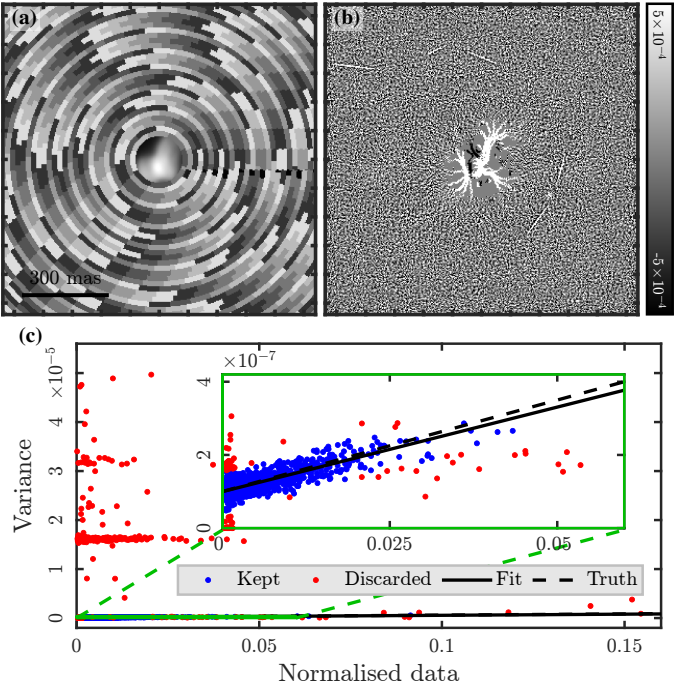
 
	\centering
        
	\newcommand{\PathFig}{Fig_App_Noise/}
        
	\newcommand{\spaceLine}{-0.05cm}
        
	\newcommand{\LineRatio}{0.975}
	\newcommand{\subfigColor}{black}		
		
	\newcommand{\FigOne}{\PathFig Res_map}  
	\newcommand{\FigTwo}{\PathFig Res_bar}  
	\sbox1{\includegraphics{\FigOne}}
	\sbox2{\includegraphics{\FigTwo}}
	
	\newcommand{\ColumnWidth}[1]
		{\dimexpr \LineRatio \linewidth * \AspectRatio{#1} / (2*\AspectRatio{1} + \AspectRatio{2}) \relax
		}
	\newcommand{\ColumnGap}{\hspace {\dimexpr \linewidth /4 - \LineRatio\linewidth /4}}
			
	\begin{tabular}{
		@{\ColumnGap}
		M{\ColumnWidth{1}}
		@{\ColumnGap}
		M{\ColumnWidth{1}}
		@{\ColumnGap}
		M{\ColumnWidth{2}}
		@{\ColumnGap}
		}		 

		\subfigimg[width=\linewidth,pos=ul,font=\fontfig{\subfigColor}]{$\;$(a)}{0.5}{\PathFig Arcs_map}
		&
		\subfigimg[width=\linewidth,pos=ul,font=\fontfig{\subfigColor}]{$\;$(b)}{0.5}{\PathFig Res_map}
		&
		\subfigimg[width=\linewidth,pos=ul,font=\fontfig{\subfigColor}]{}{0.0}{\PathFig Res_bar}
		\\[\spaceLine]
		\multicolumn{3}{c}{
			\subfigimg[width=0.95\linewidth,pos=ul,font=\fontfig{\subfigColor}]{(c)}{0.0}{\PathFig Model}
		}
	\end{tabular}
	
	\caption{\label{fig:noise_model} Empirical fit of the noise model,~$\eta\data\Paren{\Vx}+\varRON$, directly from the data. \refpan{a}:~Visualisation of the arcs centred on the main object on which the model is fitted. \refpan{b}:~Approximated map of the noise after removing a median filter on the data. \refpan{c}:~Estimated noise model (black curve) compared with the true model (dashed curve). In blue (resp. red): points that are kept (resp. discarded) by the robust fit. The green box is a zoom on the relevant points.}
\end{figure}

The attentive reader may notice that the proposed method implies knowledge of the noise model parameters,~$\paren{\eta, \varRON}$, in \refeq{eq:weight}. If they are known in the simulated case of \refsec{sec:res_sim}, they are a priori unknown for the real data of \refsec{sec:res_real}. In principle, the noise model can be obtained through proper calibration and analysis of both calibration and science data~\citep[see for example,][]{Berdeu:20_PIC, Denneulin:21_RHAPSODIE}. Nonetheless, most of the data used in this paper are ZIMPOL data from the ESO Large Programme described in \refsec{sec:res_data} that were already reduced~\citep{Vernazza:21_Large_program}. It was beyond the scope of this paper to fully recharacterise the ZIMPOL sensor noise and rerun a reduction procedure. As a consequence, I chose to implement a pragmatic approach that could empirically fit these parameters directly in the data of interest and that could be easily applied to reduced data from other instruments. The principle of the method is pictured in \reffig{fig:noise_model}.

The noise model of \refeq{eq:weight} is basically an affine law that links the data intensity and its associated variance. With the assumption that the noise parameters are constant over the FoV, \refeq{eq:weight_const}, the two noise parameters, $\paren{\eta, \varRON}$, can be fitted on the full data image using the diversity of information, as is implemented in \refalg{alg:noise}.

First, as is shown in \refsubfig{fig:noise_model}{b}, the noise map was approximated by removing from the data, a smoothed image obtained after applying a median filter of $5\times5$ pixels, \reflin{noise}{med}. The pixels were then split among $\n{arc}$ arcs (with a typical width of $5$~pixels and length of $20$~pixels for the ZIMPOL data) centred on the main body, shown in \refsubfig{fig:noise_model}{a}. The intensities (resp. variances) of the model were empirically estimated by averaging the data (resp. by computing the variance of the noise map) on each arc, \reflin{noise}{avg} (resp. \reflin{noise}{var}). To be robust with regard to the presence of outliers among this point cloud, \refsubfig{fig:noise_model}{c}, the affine law was robustly fitted via an approach based on the median absolute deviation of the model from the empirical points, \reflin{noise}{mad}, a robust estimator of the standard deviation~\citep{Huber:11_robust}.

This method was tested on the simulated data of \refsec{sec:res_sim}, for which the true noise parameters are known. The results of the fit, given in \refsubfig{fig:noise_model}{c}, show that the noise model was successfully estimated.

I point out here that these noise parameters could be iteratively refined during the iterations of the alternate algorithm once the model of the convolution between the object and the PSF, namely the theoretical intensity map, and the model of the halo residuals, namely the theoretical noise map, are better constrained.\footnote{See for example the documentation of the Epifluorescence DEconvolution MICroscopy plug-in~\citep[EpiDEMIC,][]{Soulez:12} at \href{https://icy.bioimageanalysis.org/plugin/epidemic/}{https://icy.bioimageanalysis.org/plugin/epidemic/}.} This was nonetheless beyond the scope of this paper and is kept for a future work. Indeed, the homogeneity of the weight maps in \reffigs{fig:comp_instru}{fig:Elektra_data} shows that the proposed method is a good enough approximation, correctly whitening both the readout noise and the photon shot noise.

\begin{algorithm}[h!]
\caption{\label{alg:noise} Empirical fit of the noise model.}
\begin{algorithmic}[1]
	\small  
	  
	\State $\Vdata^{\Tag{noise}} \gets \Vdata - \f{med}{\Vdata}$
		\commentalgo{Applying a median filter to the data}
		\label{lin:noise:med}

	\For{$a\text{ from } 1 \text{ to } \n{arc}$}
		\commentalgo{Loop on the arcs}
	
		\State $\Reua\Paren{a}\gets \f{avg}{\Brace{\data\Paren{\x}}_{\Vx\in \text{arc}_a}}$
			\commentalgo{Empirical average value on $a^\text{th}$ arc}
			\label{lin:noise:avg}
	
		\State $\Nguu\Paren{a}\gets \f{var}{\Brace{\data^{\Tag{noise}}\Paren{\x}}_{\Vx\in \text{arc}_a}}$
			\commentalgo{Empirical variance on $a^\text{th}$ arc}
			\label{lin:noise:var}
	\EndFor
	
	\State $\mathcal{V}^{\Tag{val}}\gets \Bbrack{1,\n{arc}}$
		\commentalgo{Initialisation: all arcs are valid}
	
	\State $r \gets \Paren{\eta, \varRON, a} \mapsto \Nguu\Paren{a}-\Paren{\eta\Reua\Paren{a}+\varRON} $
		\commentalgo{Defining the residuals}
		
	\For{$i\text{ from } 1 \text{ to } 10$}
		\commentalgo{10 robust fits}
		\label{lin:noise:mad}
	
		\State $\Paren{\eta, \varRON} = \argmin{\tilde{\eta} \geq 0, \tilde{v}_{0} \geq 0}{\sum_{a\in\mathcal{V}^{\Tag{val}}}\Abs{r\Paren{\tilde{\eta}, \tilde{v}_{0}, a}}}$
			\commentalgo{Robust linear fit}
		
		\State $\sigma = 1.4826\f{med}{\Brace{\Abs{r\Paren{\eta, \varRON, a}}}_{a\in\mathcal{V}^{\Tag{val}}}}$	
			\commentalgo{Median absolute deviation of the fit residuals}
			
	\State $\mathcal{V}^{\Tag{val}}\gets \Brace{a\in\Bbrack{1,\n{arc}}\st\Abs{r\Paren{\eta, \varRON, a}}<3\sigma}$
		\commentalgo{Updating set of valid arcs}
	\EndFor
	\State \Return $\paren{\eta, \varRON}$
\end{algorithmic}
\end{algorithm}

\section{Additional frames of \Elektra}
\label{app:Elektra_data}

\begin{figure*}[p!] 
        \centering
        
        \newcommand{\PathFig}{Fig_comp_instru/}
        
        \newcommand{\LineRatio}{0.95}
        
        \newcommand{\fontTxt}[1]{\textbf{\small #1}}
        
        \newcommand{\widthTxt}{12pt}
        
        \newcommand{\sizeTxt}[1]{\Large{#1}}
        
        \newcommand{\spaceLine}{-2pt}
        
        \newcommand{\widthFig}{\dimexpr (\linewidth - \widthTxt)}
        
        \newcommand{\subfigColor}{white}        
        
        \sbox1{\includegraphics{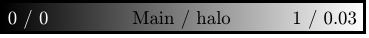}}
        
        \newcommand{\ColumnWidth}
                {\dimexpr \LineRatio \widthFig * \AspectRatio{1} / (\AspectRatio{1}*5) \relax
                }
        \newcommand{\ColumnGap}{\hspace {\dimexpr \widthFig /7 - \LineRatio\widthFig /7 }}    
        
		\newcommand{\InsertLine}[3]{
			\rotatebox[origin=l]{90}{\fontTxt{#2}}
			&
			\subfigimg[width=\linewidth,pos=ul,font=\fontfig{\subfigColor}]{#3}{0.0}{#1_Data}
			&
			\subfigimg[width=\linewidth,pos=ul,font=\fontfig{\subfigColor}]{}{0.0}{#1_Model}
			&
			\subfigimg[width=\linewidth,pos=ul,font=\fontfig{\subfigColor}]{}{0.0}{#1_PSF}
			&
			\subfigimg[width=\linewidth,pos=ul,font=\fontfig{\subfigColor}]{}{0.0}{#1_Res}
			&
			\subfigimg[width=\linewidth,pos=ul,font=\fontfig{\subfigColor}]{}{0.0}{#1_Weight}
			\\[\spaceLine]
        }

        \begin{tabular}{
                @{\ColumnGap}
                M{\widthTxt}
                @{\ColumnGap}
                M{\ColumnWidth}
                @{\ColumnGap}
                M{\ColumnWidth}
                @{\ColumnGap}
                M{\ColumnWidth}
                @{\ColumnGap}
                M{\ColumnWidth}
                @{\ColumnGap}
                M{\ColumnWidth}
                @{\ColumnGap}
                }
                
            &
            \fontTxt{Data}
            &
            \fontTxt{Halo and object models}
            &
            \fontTxt{Point spread function}
            &
            \fontTxt{Halo residuals}
            &
            \fontTxt{Robust weights}
			\\
                
			\InsertLine{\PathFig/ZIMPOL_2019-07-30_1}{2019-07-30 (1)}{}
			\InsertLine{\PathFig/ZIMPOL_2019-08-04_1}{2019-08-04 (1)}{}
			\InsertLine{\PathFig/ZIMPOL_2019-08-05_1}{2019-08-05 (1)}{}
			\InsertLine{\PathFig/ZIMPOL_2019-08-06_2}{2019-08-06 (2)}{}
			\InsertLine{\PathFig/Bar}{}{}
 			         
        \end{tabular}

        \caption{\label{fig:Elektra_data} Additional frames of \Elektra with ZIMPOL (see caption of \reffig{fig:comp_instru}).}
\end{figure*}

\reffigfull{fig:Elektra_data} comes in addition to \reffig{fig:comp_instru} and presents the second frame of the couples introduced in \reffig{fig:Elektra_profiles}. This figure supports the consistency of my method that produces similar results within each epoch, in terms of PSF structures and dynamics as well as moon enhancement.

\end{appendix}

\end{document}